\documentclass[traditabstract]{aa}
\usepackage{natbib,twoopt}
 \bibpunct{(}{)}{;}{a}{}{,}    
\usepackage{graphics}
\usepackage{txfonts}
\usepackage{color}
\usepackage{ulem}
\usepackage{times}
\usepackage{psfig}
\begin{document}
\title{Searching for metal-deficient emission-line galaxy candidates: 
the final sample of the SDSS DR12 galaxies
}
\author{N. G. \ Guseva \inst{1,2}    
\and Y. I. \ Izotov \inst{1,2}
\and K. J. \ Fricke \inst{1,3}
\and C. \ Henkel \inst{1,4}
}
\offprints{N. G. Guseva, guseva@mao.kiev.ua}
\institute{          Max-Planck-Institut f\"ur Radioastronomie, Auf dem H\"ugel 
                     69, 53121 Bonn, Germany
\and
                     Main Astronomical Observatory,
                     Ukrainian National Academy of Sciences,
                     Zabolotnoho 27, Kyiv 03680,  Ukraine
\and 
                     Institut f\"ur Astrophysik, G\"ottingen Universit\"at, 
                     Friedrich-Hund-Platz 1, 37077 G\"ottingen, Germany 
\and
                     Astronomy Department, King Abdulaziz University, 
                     P.O.Box 80203, Jeddah 21589, Saudi Arabia
}
\date{Received \hskip 2cm; Accepted}

\abstract{
We present a spectroscopic study of metal-deficient 
dwarf galaxy candidates,
selected from the SDSS DR12.
The oxygen abundances were derived using the direct method in galaxies with
the electron temperature-sensitive emission line 
[O~{\sc iii}]$\lambda$4363$\AA$\ measured 
 with an accuracy better than 30\%.
The oxygen abundances for the remaining galaxies
with larger uncertainties of the [O~{\sc iii}]$\lambda$4363$\AA$\ line
fluxes were calculated using a strong-line semi-empirical  method by 
Izotov and Thuan.
 The resulting sample consists of 287 low-metallicity candidates with 
oxygen abundances below  12 +log O/H = 7.65  including 23 extremely 
metal-deficient (XMD) candidates with 12 +log O/H $\leq$ 7.35. 
Ten out of sixteen XMDs known so far (or $\sim$ 60\%) have been
discovered by our team using the direct method. 
Three XMDs were found in the present study.
 We study relations between global parameters of low-metallicity galaxies,
including absolute optical magnitudes, H$\beta$ luminosities 
(or equivalently star formation rates), stellar masses, mid-infrared 
colours, and oxygen abundances.
 Low-metallicity and XMD galaxies strongly deviate to lower metallicities
in $L-Z$, $L$(H$\beta$)-$Z$ and $M_*$-$Z$ diagrams than in relations 
obtained for large samples of low-redshift, star-forming galaxies with
non-restricted metallicities. 
 These less chemically evolved galaxies with stellar masses 
$\approx$ 10$^6$ -- 10$^8$$M_{\odot}$, H$\beta$ luminosities $\approx$ 
10$^{38}$ -- 10$^{41}$ erg s$^{-1}$, SFR $\approx$ 0.01 -- 1.0 M$_\odot$ yr$^{-1}$, 
and sSFR $\sim$ 50 Gyr$^{-1}$ have physical conditions which may be 
characteristic of high-redshift low-mass star-forming galaxies
which are still awaiting discovery. 
}
\titlerunning{New XMD galaxies in SDSS DR12}
\authorrunning{N. G. Guseva et al.}
\maketitle

\section {Introduction}

The discovery and subsequent study of gas-phase elemental abundances and
global characteristics of low-metallicity and extremely metal-deficient 
(XMD) star-forming (SF) galaxies are important for investigations of the 
interstellar medium (ISM) conditions prevailing in the early Universe. 
They are also relevant in order to provide constraints to numerical models of
chemical evolution and to study the young massive stellar population in the 
galaxies.
 These chemically unevolved galaxies allow us to study the key astrophysical 
processes occurring in the early Universe, and they also provide measurements 
of the primordial helium abundance as a test of Big Bang nucleosynthesis models.
 It is necessary to increase statistics of these galaxies to better constrain
the lower metallicity threshold which is determined by the most metal-deficient 
galaxies. 
This observationally determined threshold near 12 + logO/H $\sim$7 confirms 
the idea of the prior enrichment of the pristine gas by 
Population III stars \citep[see e.g., ][]{T05}. 


\setcounter{table}{0}

\begin{table*}
  \caption{General characteristics of newly identified XMD candidates \label{tab1}}
\begin{tabular}{llccccccl} \hline \hline
Name&Method$^{\rm a}$&R.A.(J2000.0)$^{\rm b}$&Dec.(J2000.0)$^{\rm b}$&
redshift&$g$$^{\rm c}$&M$_g$$^{\rm d}$&M$_{\star}$$^{\rm e}$&
Other names \\  \hline
J0015$+$0104  &E& 00:15:20.68 &   01:04:36.99  & 0.0069$\pm$0.0002 & 17.91$\pm$0.03 & $-$14.70$\pm$0.04 &  5.234 & \\
J0042$+$3247  &E& 00:42:33.37 &   32:47:21.01  & 0.1426$\pm$0.0003 & 22.13$\pm$0.08 & $-$17.10$\pm$0.09 &  7.883 & \\
J0100$-$0028  &E& 01:00:56.93 & $-$00:28:43.90~~~ & 0.0192$\pm$0.0000 & 20.85$\pm$0.05 & $-$14.43$\pm$0.05 &  6.965 & \\
J0122$+$0048  &D& 01:22:41.61 &   00:48:42.00  & 0.0574$\pm$0.0000 & 21.62$\pm$0.06 & $-$16.16$\pm$0.06 &  6.593 & \\
J0137$+$1810  &D:$^{\rm f}$& 01:37:54.44 & 18:10:35.98 & 0.0659$\pm$0.0000 & 22.28$\pm$0.17 & $-$16.08$\pm$0.17 & 6.289 & \\
J0141$+$2124  &D:& 01:41:33.22 &   21:24:50.33  & 0.1638$\pm$0.0002 & 22.53$\pm$0.19 & $-$17.72$\pm$0.19 &  7.220 & \\ 
J0143$+$1958  &E& 01:43:15.15 &   19:58:06.10  & 0.0017$\pm$0.0000 & 21.76$\pm$0.06 & ~\,$-$7.66$\pm$0.06 &  4.205 & \\
J0153$+$0104  &D& 01:53:11.96 &   01:04:40.10  & 0.0632$\pm$0.0000 & 21.51$\pm$0.06 & $-$16.23$\pm$0.06 &  6.567 & \\
J0207$-$0821  &E& 02:07:24.77 & $-$08:21:43.60~~~  & 0.0127$\pm$0.0000 & 20.23$\pm$0.03 & $-$14.46$\pm$0.03 &  6.635 & \\
J0223$-$0918  &E& 02:23:02.68 & $-$09:18:22.40~~~  & 0.0503$\pm$0.0000 & 20.99$\pm$0.06 & $-$15.84$\pm$0.06 &  7.940 & \\
J0739$+$4434  &D:& 07:39:02.59 &   44:34:26.18  & 0.0464$\pm$0.0000 & 22.01$\pm$0.09 & $-$14.64$\pm$0.09 &  5.798 & \\
J0757$+$1423  &E& 07:57:01.03 &   14:23:47.68  & 0.0012$\pm$0.0000 & 15.52$\pm$0.01 & $-$13.33$\pm$0.01 & 5.912 & UGC 04115 \\ 
J0945$+$3835  &D:& 09:45:19.55 &   38:35:52.90  & 0.0723$\pm$0.0000 & 21.78$\pm$0.06 & $-$16.82$\pm$0.06 &  7.234 & \\
J0955$+$6442  &E& 09:55:31.45 &   64:42:50.06  & 0.0032$\pm$0.0000 & 17.91$\pm$0.01 & $-$12.85$\pm$0.01 &  6.176 & \\
J0959$+$4626  &E& 09:59:05.76 &   46:26:50.49  & 0.0020$\pm$0.0000 & 17.80$\pm$0.01 & $-$12.20$\pm$0.01 &  6.272 & \\
J1000$+$3032  &E& 10:00:36.54 &   30:32:09.78  & 0.0017$\pm$0.0000 & 17.77$\pm$0.01 & $-$11.76$\pm$0.01 &  6.421 & \\
J1034$+$1546  &E& 10:34:05.40 &   15:46:50.14  & 0.0041$\pm$0.0000 & 18.01$\pm$0.01 & $-$13.43$\pm$0.01 &  7.236 & \\
J1036$+$2036  &E& 10:36:39.47 &   20:36:15.80  & 0.0549$\pm$0.0000 & 21.59$\pm$0.06 & $-$15.61$\pm$0.06 &  6.982 & \\
J1119$+$0935  &E& 11:19:28.09 &   09:35:44.28  & 0.0036$\pm$0.0000 & 17.13$\pm$0.01 & $-$14.02$\pm$0.01&  7.154 & \\ 
J1208$+$3727  &E& 12:08:09.75 &   37:27:24.65  & 0.0036$\pm$0.0000 & 17.76$\pm$0.02 & $-$14.06$\pm$0.02 &  6.825 & \\
J1220$+$4915  &D& 12:20:51.61 &   49:15:55.48  & 0.0123$\pm$0.0000 & 20.89$\pm$0.04 & $-$12.94$\pm$0.04 &  5.315 & \\ 
J1228$-$0125  &E& 12:28:45.54 & $-$01:25:26.90~~~  & 0.0728$\pm$0.0000 & 22.09$\pm$0.10 & $-$15.58$\pm$0.10 &  7.172 & \\
J1244$+$3212  &E& 12:44:11.17 &   32:12:21.69  & 0.0022$\pm$0.0000 & 15.15$\pm$0.01 & $-$15.72$\pm$0.01 &  6.787 &  NGC 4656 \\
J1258$+$1412  &E& 12:58:40.13 &   14:12:28.80  & 0.0007$\pm$0.0000 & 21.60$\pm$0.05  & ~\,$-$6.04$\pm$0.05 & 3.658&  \\ 
J1258$+$1413  &E& 12:58:40.20 &   14:13:00.79  & 0.0007$\pm$0.0000 & 15.98$\pm$0.01 & $-$12.25$\pm$0.01 &  6.535 &  UGC 08091 \\
J1308$+$2002  &E& 13:08:28.41 &   20:02:01.93  & 0.0049$\pm$0.0000 & 17.54$\pm$0.01 & $-$14.12$\pm$0.01 &  7.263 & \\ 
J1320$+$1338  &E& 13:20:47.80 &   13:38:04.24  & 0.0227$\pm$0.0000 & 17.90$\pm$0.01 & $-$17.27$\pm$0.01 & 8.735 & AGC 233628 \\ 
J1444$+$4237  &D:& 14:44:12.80 &   42:37:44.01  & 0.0021$\pm$0.0000 & 15.86$\pm$0.05 & $-$14.40$\pm$0.05 &  7.075 & UGC 09497 \\ \hline
\end{tabular}

$^{\rm a}$The coding is as follows: E -- semi-empirical method by \citet{IT07}, D -- direct method using the temperature-sensitive ratio of emission lines [O {\sc iii}]$\lambda$4363/($\lambda$4959+$\lambda$5007). \\
$^{\rm b}$Equatorial coordinates. \\
$^{\rm c}$SDSS $g$ magnitude. \\
$^{\rm d}$Absolute $g$ magnitude. \\
$^{\rm e}$Log of stellar mass in solar masses, typical error is $\pm$0.2 dex. \\
$^{\rm f}$galaxies with an [O~{\sc iii}]$\lambda$4363$\AA$\
signal-to-noise emission line ratio between 2 and 3 are marked by a colon. 

  \end{table*}


\setcounter{figure}{0}
\begin{figure*}
\hbox{
\psfig{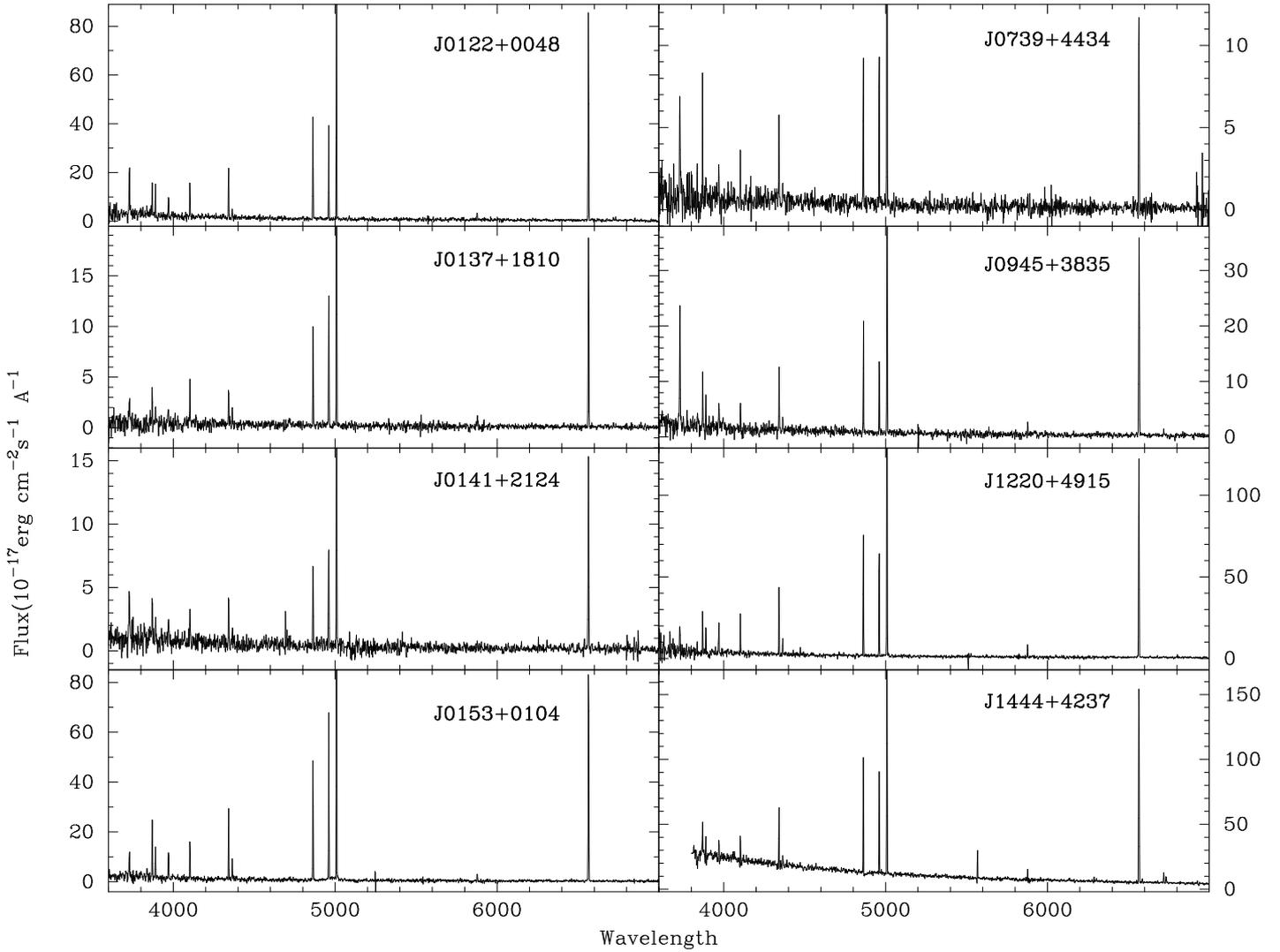}
}
\caption{ SDSS spectra of the nine XMD candidates for which abundances
were obtained by direct methods. Five galaxies (J0137$+$1810, J0141$+$2124,
J0739$+$4434, J0945$+$3835, J1444$+$4237) have signal-to-noise ratios 
of 2$\sigma$ $<$ S/N $<$ 3$\sigma$ in the 
[O {\sc iii}]$\lambda$4363$\AA$\ emission line. 
}
\label{fig1}
\end{figure*}

\setcounter{figure}{1}
\begin{figure*}
\hbox{
\psfig{figure=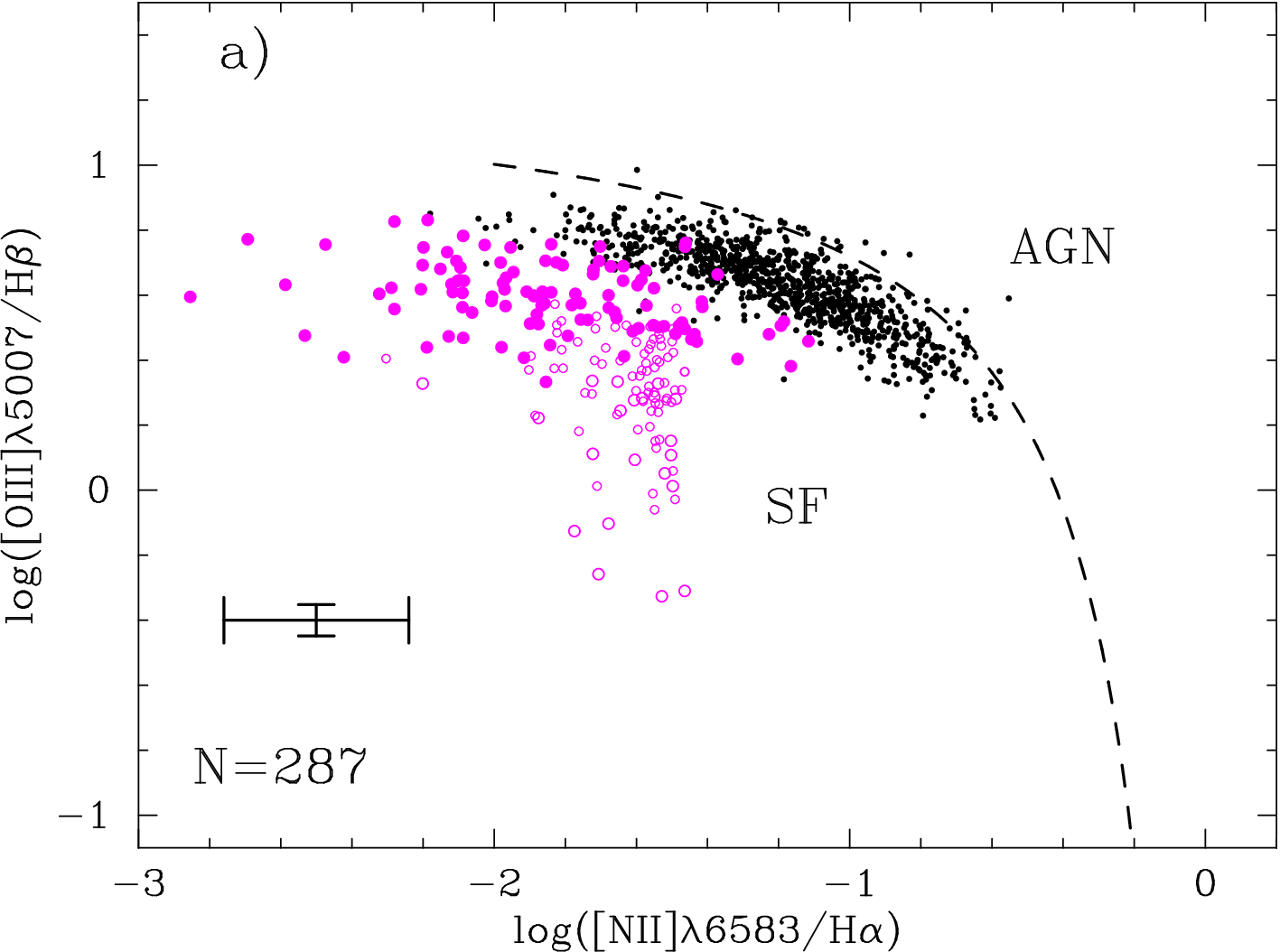,angle=-90,width=8.7cm,clip=}
\psfig{figure=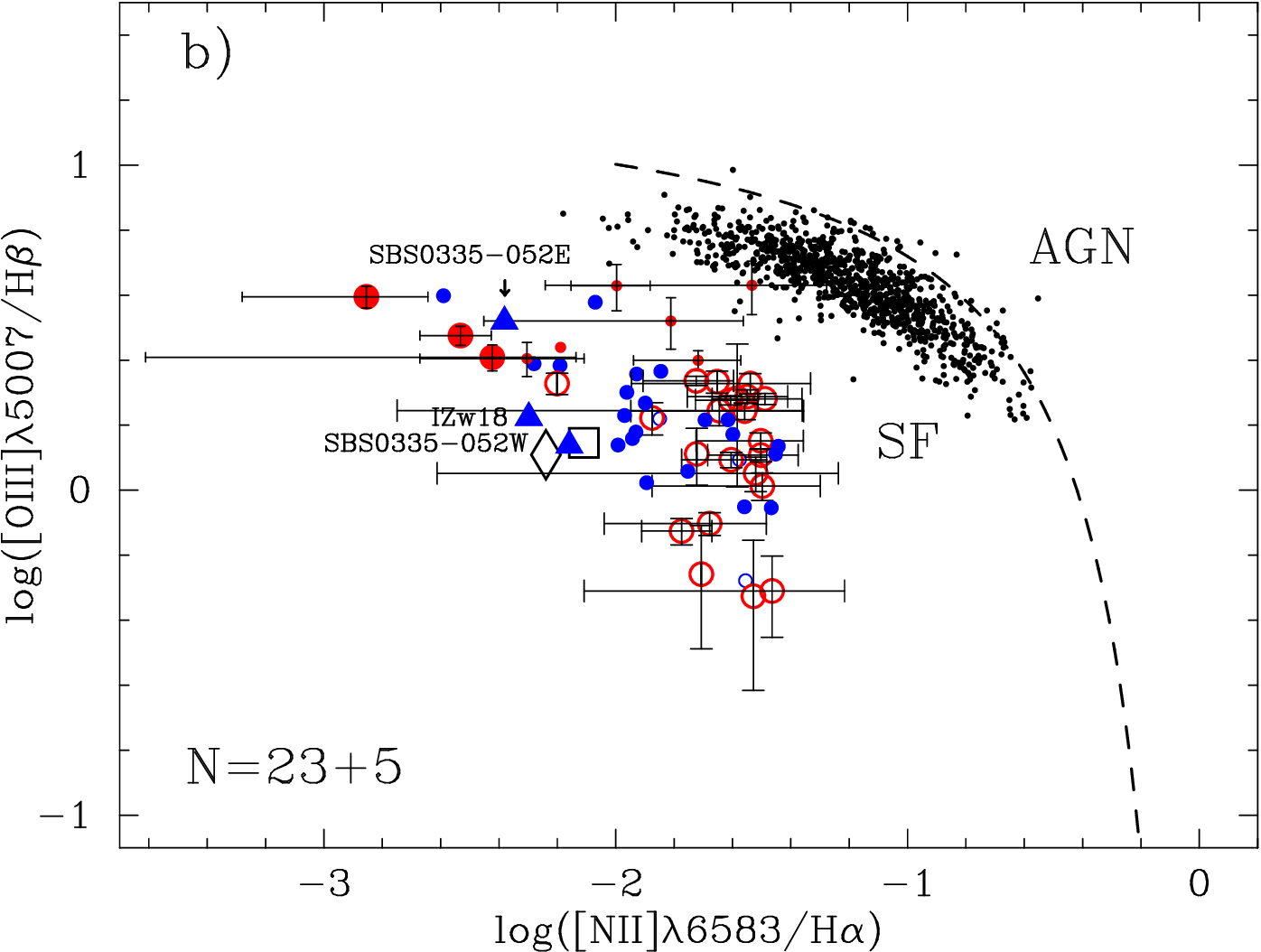,angle=-90,width=8.7cm,clip=}
}
\caption{log([O{\sc iii}]$\lambda$5007/H$\beta$) vs. 
log([N{\sc ii}]$\lambda$6583/H$\alpha$) diagnostic 
diagram \citep[BPT, ][]{BPT81}.
{\bf a)} Low-metallicity candidates with 12 + logO/H $\leq$ 7.65.
Magenta filled circles denote the galaxies with oxygen abundances derived 
using the direct method, 
and magenta open circles indicate galaxies with 
oxygen abundances derived with the semi-empirical method by \citet{IT07}. 
  Error bars at the lower left corner represent the mean range of errors
for individual galaxies.
{\bf b)} XMD candidates with 12 + logO/H $\leq$ 7.35. XMD candidates 
with metallicities derived by the direct and semi-empirical methods
are denoted by red filled circles and red open circles, respectively.
   The galaxies with the 
[O {\sc iii}]$\lambda$4363$\AA$\ emission lines detected at a  
level higher than 3$\sigma$ are denoted by large red filled symbols and 
galaxies detected at 2$\sigma$ -- 3$\sigma$ levels by small red filled circles.
We also show the XMD emission-line galaxies and candidates 
with 12 + logO/H $\leq$ 7.35 by \citet{IT07}, \citet{I12} and 
\citet{Guseva2015} by small blue circles.  The well-known metal-deficient BCDs 
I Zw 18, SBS 0335--052W, and SBS 0335--052E are represented by filled blue 
triangles and are labelled. 
The extremely metal-poor dwarf galaxies Leo P \citep{Skillman2013} 
 and AGC 198691 \citep{Hirschauer2016} are plotted by a large open black 
square and a large open black diamond, respectively.
  For comparison we also plot in a) and b) a sample of 803 
luminous compact galaxies \citep[LCGs, black dots, ][]{I11}.
 The dashed line from \citet{K03}  
separates star-forming galaxies (SF) from active galactic nuclei (AGN). 
}
\label{fig2}
\end{figure*}


  Much attention has been focused on searching for XMD galaxies resulting in
their steadily increasing number \citep{P05,Pustilnik2011,I06b,I12,IT07,IT09,G07,Guseva2015,B12,Hirschauer2016,Almeida2016}.
The metallicity of a galaxy is customarily expressed in terms of the 
oxygen abundance 12 + log O/H.
  Exact values of the metallicity defining extremely metal-poor (XMP)
and XMD galaxies are different in different studies
\citep[e.g., ][]{KO2000,IT07,I12,Guseva2015,Almeida2016}, ranging from 
$\sim$7.3 to $\sim$7.8.
 In this paper we continue the search for the least metal-abundant galaxies 
in the local Universe, extracting them from the SDSS surveys.
 Our metallicity-restricted samples consist of a low-metallicity sample with
oxygen abundance 12 +logO/H $\leq$ 7.65 including a subsample of XMD galaxies 
with 12 +logO/H $\leq$ 7.35.

 The paper is organised as follows. 
 The sample and element abundance determination are described in 
Sect. \ref{S2}.  
 The results are presented in Sect. \ref{S3}. 
 More specifically,
the global properties of low-metallicity and XMD emission-line galaxies
are analysed in Sect. \ref{S3s1}, the luminosity-metallicity
and mass-metallicity relations are discussed in Sect. \ref{S3s2}.
 Relations between mid-infrared $WISE$ colours and global parameters of the 
low-metallicity sample are presented in Sect. \ref{S3s3}.
 Our main conclusions are summarised in Sect. \ref{S4}.

\setcounter{figure}{2}

\begin{figure*}
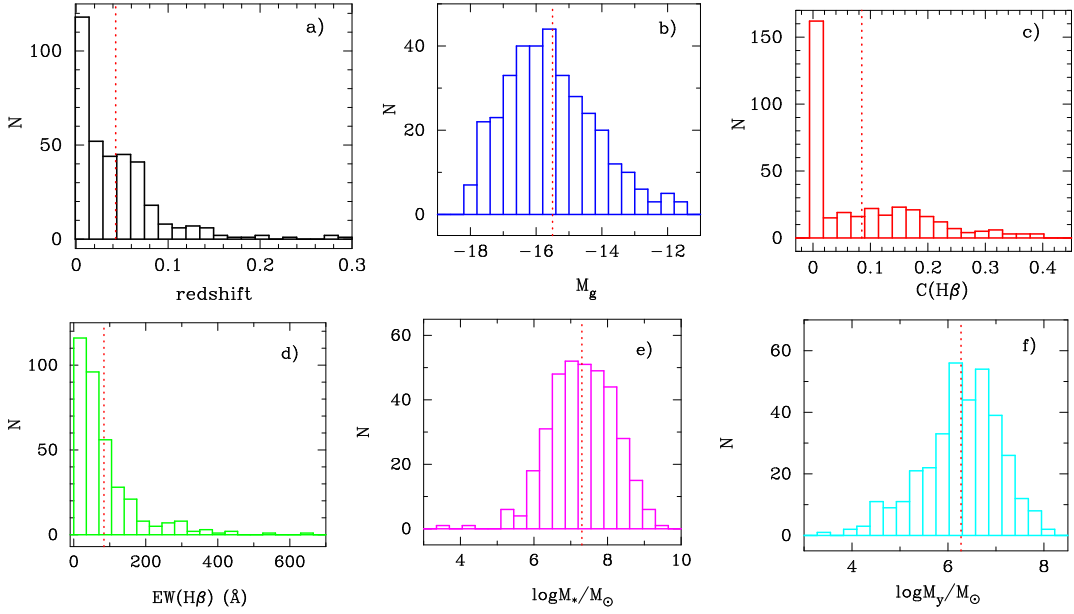

\vspace{0.05cm}
\hbox{
\hspace{1.50cm}\psfig{figure=fig3a.ps,angle=-90,width=4.7cm,clip=}
\hspace{0.15cm}\psfig{figure=fig3b.ps,angle=-90,width=4.2cm,clip=}
\hspace{0.3cm}\psfig{figure=fig3c.ps,angle=-90,width=4.5cm,clip=}
}
\vspace{0.2cm}
\hbox{
\hspace{1.50cm}\psfig{figure=fig3d.ps,angle=-90,width=4.2cm,clip=}
\hspace{0.3cm}\psfig{figure=fig3e.ps,angle=-90,width=4.4cm,clip=}
\hspace{0.3cm}\psfig{figure=fig3f.ps,angle=-90,width=4.6cm,clip=}
}
\caption{Distributions of low-metallicity candidates with 
12 + logO/H $\leq$ 7.65
(N=287, the same objects as those shown in Fig. \ref{fig2}a by
magenta symbols) on 
{\bf a)} 
the redshift $z$; {\bf b)} the extinction-corrected absolute
SDSS $g$-magnitude $M_g$; {\bf c)} the extinction coefficient 
$C$(H$\beta$); {\bf d)} the equivalent width EW(H$\beta$) of the H$\beta$
emission line; {\bf e)} the total stellar mass $M_\star$; 
{\bf f)} the mass $M_y$ of the young stellar population.
Dotted vertical lines in all panels indicate mean values of the distributions.}
\label{fig3}
\end{figure*}

\setcounter{table}{3}

\begin{table}
  \caption{List of the most metal-deficient emission-line galaxies 
 \label{tab4}}
\begin{tabular}{llccc} \hline \hline
Name& Method$^{\rm a}$& O/H$^{\rm b}$ & Ref.$^{\rm c}$& Other \\  
&&&&names \\ 
\hline
J0015$+$0104  &E&   7.17 $\pm$ 0.03 & 1 & \\
J0042$+$3247  &E&   7.32 $\pm$ 0.04 & 1 & \\
J0100$-$0028  &E&   7.05 $\pm$ 0.18 & 1 & \\
J0113$+$0052  &D&   7.15-7.32& 2,3 & \\
J0122$+$0048  &D&   7.20 $\pm$ 0.17 & 1 & \\
J0137$+$1810  &D:$^{\rm d}$&   7.32 $\pm$ 0.22  & 1 & \\
J0141$+$2124  &D:&   7.35 $\pm$ 0.30 & 1 & \\
J0143$+$1958  &E&   7.29 $\pm$ 0.10 & 1 & \\
J0153$+$0104  &D&   7.35 $\pm$ 0.10 & 1 & \\
J0207$-$0821  &E&   7.35 $\pm$ 0.05 & 1 & \\
J0223$-$0918  &E&   7.33 $\pm$ 0.11 & 1 & \\
J0254$+$0035  &E&   7.28 $\pm$ 0.05 & 5 & \\
J0337$-$0502  &D&  6.86-7.22 & 4 & SBS0335$-$052W \\
J0337$-$0502  &D&  7.11-7.32 & 4 & SBS0335$-$052E \\
J0739$+$4434  &D:&   7.34 $\pm$ 0.28  & 1 & \\
J0757$+$1423  &E&   7.13 $\pm$ 0.09 & 1 & UGC04115 \\
J0812$+$4836  &E&   7.28 $\pm$ 0.06 & 5 & \\
J0834$+$5905  &D&   7.24 $\pm$ 0.08 & 2 & \\
J0926$+$3343  &D&   7.12 $\pm$ 0.02 & 11 & \\
J0934$+$5514  &D&   7.17-7.22&  6 & IZw18 \\
J0943$+$3325  &D&   7.02 $\pm$ 0.03 & 13 & AGC198691 \\
J0945$+$3835  &D:&   7.33 $\pm$ 0.38 & 1 & \\ 
J0955$+$6442  &E&   7.33 $\pm$ 0.02 & 1 & \\
J0956$+$2849  &E&   7.14-7.16&  2,5,7 & DDO68 \\
J0959$+$3044  &D&   7.30 $\pm$ 0.05 & 8 & LeoA \\
J0959$+$4626  &E&   7.22 $\pm$ 0.02 & 1 & \\
J1000$+$3032  &E&   7.35 $\pm$ 0.04 & 1 & \\
J1021$+$1805  &D&   7.17 $\pm$ 0.04 & 12 & LeoP \\
J1034$+$1546  &E&   7.33 $\pm$ 0.02 & 1 & \\
J1036$+$2036  &E&   7.35 $\pm$ 0.04 & 1 & \\
J1056$+$3608  &D&   7.16-7.32&  2 & \\
J1057$+$1358  &E&   7.18 $\pm$ 0.07 & 2 & \\
J1119$+$0935  &E&   7.27 $\pm$ 0.02 & 1 & \\
J1121$+$3744  &E&   7.12 $\pm$ 0.06 & 2 & \\
J1146$+$4050  &E&   7.34 $\pm$ 0.06 & 2 & \\
J1208$+$3727  &E&   7.32 $\pm$ 0.03 & 1 & \\
J1220$+$4915  &D&   7.20 $\pm$ 0.10 & 1 & \\
J1228$-$0125  &E&   7.33 $\pm$ 0.06 & 1 & \\
J1238$-$3246  &D&   7.30 $\pm$ 0.05 & 9 & UGCA292 \\
J1244$+$3212  &E&   7.28 $\pm$ 0.02 & 1 & NGC4656 \\
J1258$+$1412  &E&   7.35 $\pm$ 0.16 & 1 & \\
J1258$+$1413  &E&   7.32 $\pm$ 0.02 & 1 & UGC08091 \\
J1308$+$2002  &E&   7.22 $\pm$ 0.05 & 1 & \\
J1320$+$1338  &E&   7.32 $\pm$ 0.02 & 1 & AGC233628 \\
J1414$-$0208  &D&   7.32 $\pm$ 0.05 & 10 & \\
J1444$+$4237  &D:&   7.32 $\pm$ 0.19 & 1 & UGC09497 \\
J2104$-$0035  &D&   7.26 $\pm$ 0.03 & 3 & \\ \hline
\end{tabular}

$^{\rm a}$The coding is as follows: E -- semi-empirical method, D -- direct method. \\
$^{\rm b}$Oxygen abundance in units of 12 +logO/H. \\
$^{\rm c}${\bf References.} (1) this paper; (2) \citet{I12}; (3) \citet{I06b}; 
(4) \citet{I09}; 
(5) \citet{IT07};  (6) \citet{TI2005}; (7) \citet{IT09}; (8) \citet{S89}; 
(9) \citet{vZee2000}; (10) \citet{G07};  (11) \citet{Pustilnik2010};
(12) \citet{Skillman2013}; (13) \citet{Hirschauer2016} \\
$^{\rm d}$galaxies with signal-to-noise between 2$\sigma$ and
3$\sigma$ in [O~{\sc iii}]$\lambda$4363$\AA$\ emission line are marked by 
colons.

\end{table}

\section {Sample \label{S2}}

 We use a sample of $\sim$35190 star-forming galaxies
selected from the  SDSS DR12 spectroscopic database 
\citep[$\sim$1.5 million galaxies, ][]{A15} which 
were not included in the DR10 and earlier SDSS releases.

  We excluded galaxies with evidence of AGN activity, such as
broad emission lines or strong high-ionisation lines 
([Ne~{\sc v}]$\lambda$3426 and He {\sc ii} $\lambda$4686). We also excluded    
H~{\sc ii} regions in spiral galaxies by inspection of SDSS images.
Furthermore, high-luminosity galaxies which are 
predominantly distant galaxies with $M_g$ $<$ $-$18 mag were also excluded.
 Finally, we select only low-metallicity galaxies with gas-phase metallicity 
lower than 12 + logO/H $\leq$ 7.65 (LMG sample) and extremely metal-deficient  
galaxies with 12 + logO/H $\leq$ 7.35 (XMD sample).
  The resulting sample consists of 287 
LMG candidates including 23 XMD candidates. 
These new XMD candidates complement previous findings.
The electron temperatures and oxygen abundances for 3 out of the 23 
XMD galaxies were obtained by a direct method and are thus the most 
reliable ones. 
 
The galaxy names, method of oxygen abundance determination,
their coordinates, redshifts, apparent SDSS $g$ magnitudes, absolute $g$ 
magnitudes, and stellar masses are collected in Table~\ref{tab1}. 
 Galaxies with [O~{\sc iii}]$\lambda$4363$\AA$\
emission line signal-to-noise ratios between 2$\sigma$ and 3$\sigma$ are 
also included in the table and are marked by colons in column (2). 
  Figure~\ref{fig1} shows SDSS spectra of XMD galaxies for which 
abundances were obtained by the direct method. Five of these galaxies 
(J0137$+$1810, J0141$+$2124, J0739$+$4434, J0945$+$3835, J1444$+$4237) 
have relatively noisy spectra with 2$\sigma$ $<$ S/N $<$ 3$\sigma$ in the
[O {\sc iii}]$\lambda$4363$\AA$\ emission line.

   We measured the line fluxes and their errors using the 
IRAF\footnote{IRAF is the Image 
Reduction and Analysis Facility distributed by the National Optical Astronomy 
Observatory, which is operated by the Association of Universities for Research 
in Astronomy (AURA) under cooperative agreement with the National Science 
Foundation (NSF).} SPLOT routine.
The uncertainties of the line fluxes are propagated to calculate the 
uncertainties of electron temperatures, ionic and total heavy element 
abundances. 
 We derived the internal extinction using the decrement of
several Balmer hydrogen emission lines after correcting the observed spectra 
for the Milky Way extinction.
The line fluxes were corrected for reddening \citep{Cardelli1989}
and for underlying hydrogen stellar absorption by the application of an 
iterative procedure \citep{ITL94}.
  The observed line fluxes $F$($\lambda$)/$F$(H$\beta$),   
the extinction-corrected line fluxes  $I$($\lambda$)/$I$(H$\beta$) 
multiplied by 100, 
and the equivalent widths of emission lines for XMD candidates (direct
method) are shown in Table \ref{tab2}
(available only in the online version of the paper) together with extinction 
coefficients $C$(H$\beta$), observed H$\beta$ fluxes, and equivalent widths 
EW(abs) of hydrogen absorption stellar lines.

  The [O {\sc iii}]$\lambda$4363$\AA$\ emission lines detected at a
level  higher than 3$\sigma$ were measured in 120 low-metallicity galaxies
(LMG sample) including three XMDs (filled symbols in Figs. 2, 4, and 5).
This allows for a reliable oxygen abundance determination 
using the direct $T_{\rm e}$ method 
following the prescriptions of \citet{ITL94,ITL97} and \citet{TIL95}. 
The electron temperature $T_{\rm e}$(O {\sc iii}) was determined from the 
emission-line ratio [O {\sc iii}]$\lambda$4363/($\lambda$4959+$\lambda$5007)
following  \citet{ITL94,ITL97}.
 A semi-empirical method described by \citet{IT07} to derive 
$T_{\rm e}$(O {\sc iii})  was used for the remaining 168 
LMG and XMD candidates. 
We adopt a two-zone photoionised H {\sc ii}
region model: a high-ionisation zone with temperature $T_{\rm e}$(O {\sc iii}) 
where [O {\sc iii}] and [Ne {\sc iii}] lines originate, and a 
low-ionisation zone with temperature $T_{\rm e}$(O {\sc ii}) where [O {\sc ii}]
lines originate.
   For $T_{\rm e}$(O {\sc ii}), we use
the relation between the electron temperatures $T_{\rm e}$(O {\sc iii}) and
$T_{\rm e}$(O {\sc ii}) obtained by \citet{I06} from
the H {\sc ii} photoionisation models of \citet{SI03}. 
 The [S {\sc ii}]$\lambda$$\lambda$6717,6731$\AA$\ doublet 
is used to derive the electron density. 
Generally, the electron number
density in the H {\sc ii} regions of dwarf star-forming galaxies is low,
satisfying the condition of the low-density limit 
(e.g. $N_e$ $<$ 10$^4$ cm$^{-3}$ for the [S {\sc ii}] and [O {\sc ii}] 
emission lines and
$N_e$ $<$ 10$^5$ cm$^{-3}$ for the [O {\sc iii}] emission lines). Therefore, 
for clarity $N_e$ $=$ 100 cm$^{-3}$ was adopted in the cases of 
weak or absent of [S {\sc ii}] emission lines.
  We note, that in the low-density limit the element abundance determination 
is insensitive to the value of the electron number density and its error.

  We derived ionic and total oxygen and neon abundances  
using expressions for ionic abundances 
and the ionisation correction factors for neon by \citet{I06}. 
  In Table \ref{tab3} (available only in the online version)
the electron temperatures $T_{\rm e}$(O {\sc iii}) and $T_{\rm e}$(O {\sc ii}),
electron densities $N_e$, 
and oxygen and neon abundances are given for XMD galaxies and XMD galaxy 
candidates,
where the abundances were obtained by the direct method.
 Apart from the three XMDs with accuracy of $I$(4363) better than
3$\sigma$  we also show the five XMD galaxies that have lower levels of 
accuracy (30\% $<$ $\sigma$[$I$(4363)]/$I$(4363) $<$ 50\%). 
  Three XMD candidates have uncertainties in electron temperatures 
approximately 
equal to or better than 3$\sigma$. For the remaining five XMDs uncertainties in 
electron temperatures are in the range of 1.4$\sigma$ - 2.7$\sigma$.
  They are marked by colons in Tables \ref{tab1} and \ref{tab4} 
and denoted by small red symbols in the diagnostic diagram (Fig.2b).
 
 Uncertainties in oxygen abundances of 0.3 -- 0.4 dex for three out of the 
eight XMD galaxies in Table \ref{tab3} are somewhat high.
 Nevertheless, we keep these galaxies in the table because galaxies with 
extremely low metallicities are very rare 
in the local Universe. These data can provide useful guidelines for  
future higher sensitivity observations of the 
candidates with larger telescopes to improve the oxygen abundance 
determinations.

\section{Results \label{S3}}

\subsection{Global parameters of the low-metallicity sample \label{S3s1}}

 We show in Fig. \ref{fig2} the location of the LMG and XMD samples 
on the standard emission-line ratio diagnostic diagram \citep[BPT, ][]{BPT81}
to check the validity of our selection criteria. 
  It is seen that all newly discovered LMGs are located below the 
dashed demarcation line whose location was determined by \citet{K03}
in the region of star-forming galaxies. 
Furthermore, in agreement with results by \citet{Guseva2015} and \citet{I12}, 
the most metal-deficient galaxies in Fig. \ref{fig2}b occupy the region 
considerably below the lines separating star-forming 
galaxies and AGN and are located below the region, where the
luminous compact galaxies (LCGs) from \citet{I11} are found (black dots).

Initially the XMD region was only populated by a few galaxies, for example
I Zw 18 and SBS 0335-052. New discoveries of XMDs, for example
Leo P \citep{Skillman2013},  DDO 68 
\citep{P05,IT07,IT09}, new XMD candidates \citep{IT07,I12,Guseva2015}, and
AGC 198691 \citep{Hirschauer2016}, 
greatly increased their number (Fig. \ref{fig2}b).
Our new galaxies discussed in this paper further increase the number of 
known XMD galaxies and XMD candidates.  

However, the fraction of XMDs among emission-line galaxies
is very small. Twenty-three XMD candidates
out of $\sim$ 35190 emission-line
galaxies extracted from the SDSS DR12 
($<$ 0.1\%) have been found, indicating that these objects 
are very scarce in the Local Universe. 

  Among the known XMD galaxies (Table \ref{tab4}), none 
with an oxygen abundance 12 + logO/H $<$ 7.0 has been  
found so far, in agreement with previous results by 
\citet{IT07,I12,Guseva2015,Almeida2016};  and \citet{Hirschauer2016}.
  This threshold is very close to the neutral gas metallicity $\sim$7.0 in
some dwarf star-forming galaxies and to the metallicities in Ly$\alpha$ 
absorbers \citep{P03} supporting the idea of the prior enrichment of 
the primordial gas to the level of 12 + logO/H $\sim$7.0 
(or $\sim$ 1--2\% of Sun)
by Population III stars \citep[see e.g., ][]{T05}.
Theoretically, the existence of nearby zero-metal galaxies is not 
excluded, but they cannot be easily identified because of a lack of 
[O {\sc iii}] and [O {\sc ii}] emission lines. We also note, that such objects 
are likely much less common even than XMDs.

\setcounter{figure}{3}

\begin{figure*}
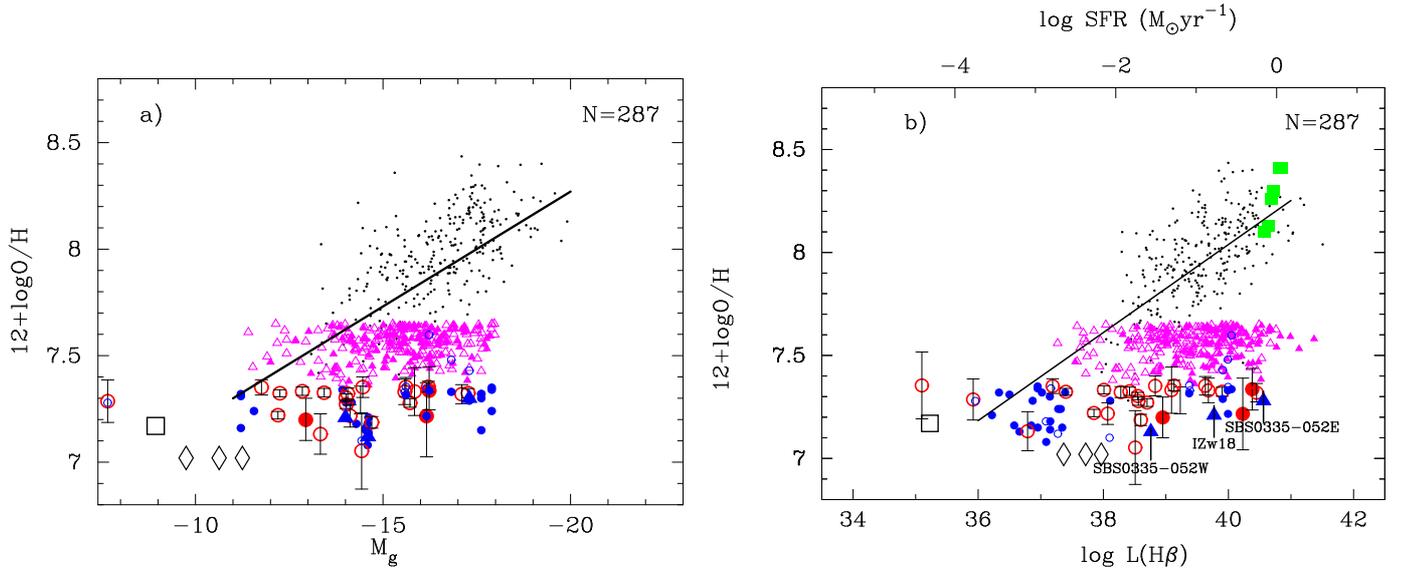


\hspace*{0.25cm}\psfig{figure=fig4a.ps,angle=-90,width=8.9cm,clip=}
\hspace*{0.25cm}\psfig{figure=fig4b.ps,angle=-90,width=8.9cm,clip=}
\caption{({\bf a}) Absolute SDSS $g$ magnitude - oxygen abundance 
relation for the galaxies with oxygen abundances 
7.35 $\leq$12 + logO/H $\leq$ 7.65 derived by the direct
method (magenta filled triangles) and for galaxies with 
oxygen abundances derived by the semi-empirical method of \citet{IT07} 
(magenta open triangles). 
   Three XMD candidates  with 12 + logO/H $\leq$ 7.35 (this paper), denoted 
by large red filled circles and identified by the direct metallicity 
determination, are shown together with large red open circles  representing 
sources with semi-empirical determinations.  
 The total number of the LMG sample shown by magenta and red symbols is 287.
  We also show the XMD emission-line galaxies and 
candidates with 12 + logO/H $\leq$ 7.35 from \citet{IT07}, \citet{I12} and 
\citet{Guseva2015} by small blue circles.
  Well-known metal-deficient BCDs 
I Zw 18, SBS 0335--052W, and SBS 0335--052E are shown by filled blue 
triangles and are labelled in b). 
 The extremely metal-poor dwarf galaxies Leo P \citep{Skillman2013} 
and AGC 198691 \citep{Hirschauer2016} are plotted respectively by 
a large open black 
square and large open black diamonds corresponding to three 
different distances.
   For comparison, the 
standard $L$ - $Z$ relation for emission-line galaxies \citep{I06,IT07}
is shown by black dots with the linear regression to these data (solid line).
({\bf b}) Extinction- and aperture-corrected H$\beta$ 
luminosity - oxygen abundance 
relation. The samples and symbols are the same as in (a). 
 The black dots represent the SDSS galaxies from \citet{I06}, and
all galaxies from \citet{IT07} and \citet{I12}. The black solid 
line  denotes the linear regression
to these data, excluding I Zw 18 and SBS 0335--052W. 
   The filled green squares represent the high-redshift star-forming 
galaxies by \citet{Cullen2014}.
}
\label{fig4}
\end{figure*}

  Fig. \ref{fig3} presents the distributions of some global characteristics of 
LMGs. 
 Average values in all six panels are depicted by red dotted vertical lines.
 The redshift distribution with the
average redshift of  $z$ $\sim$0.04 is shown in Fig. \ref{fig3}a. 
 The average redshift is approximately the same as that for the large 
non-metallicity-limited sample of $\sim$14 000 dwarf star-forming galaxies from
the SDSS DR7 \citep{I2014}.	

The brightest galaxies in our luminosity-limited sample have 
$M_g$ = $-$18 mag.
Therefore the average extinction-corrected absolute magnitude $M_g$ of $-$15.5 mag
in Fig. \ref{fig3}b is much fainter than for the sample
of compact dwarf star-forming galaxies discussed by \citet{I2014}.

  In Fig. \ref{fig3}e the galaxies are distributed in a relatively narrow 
range of aperture-corrected logarithms of total stellar masses 
log$M_\star$/$M_\odot$ from $\sim$ 6 to 9 with an average value of 7.3, which 
is lower than the value of $\sim$ 9 for the compact dwarf star-forming galaxies 
considered by \citet{I2014}.

 The average mass of the young stellar populations with ages of a few Myrs
is 6.25 in log scale (Fig. \ref{fig3}f) corresponding 
to the very high average fraction
(by mass) of $\sim$9\% of the young stellar population.
This is much higher than the average value of $\sim$ 2\% for the mass fraction
of the young stellar population in the sample of \citet{I2014}.

Galaxies from our sample are characterised by an average equivalent width 
of the H$\beta$ emission line of $\sim$80$\AA$\ (Fig. \ref{fig3}d), higher 
than the average EW(H$\beta$) $\sim$ 31$\AA$\ for the galaxies by \citet{I2014}.
 
Usually the extinction coefficient $C$(H$\beta$) at the H$\beta$ wavelength 
for dwarf star-forming  galaxies is low with an average value $\sim$0.2--0.4 
\citep[see e.g. ][]{G09,Guseva2011,Guseva2013,I2014,Guseva2015}. 
  Fig. \ref{fig3}c shows the distribution of $C$(H$\beta$) for our galaxies
varying in the range from 0 to 0.4 with very low average value 
$<$0.1 corresponding to an extinction $A_V$ in the $V$ band of less than 
0.2 mag.
  This average value is even lower than that 
for the non-metallicity-limited samples, hinting to a possibility of a light 
dependence of $C$(H$\beta$) on metallicity. 

\setcounter{figure}{4}

\begin{figure}
\psfig{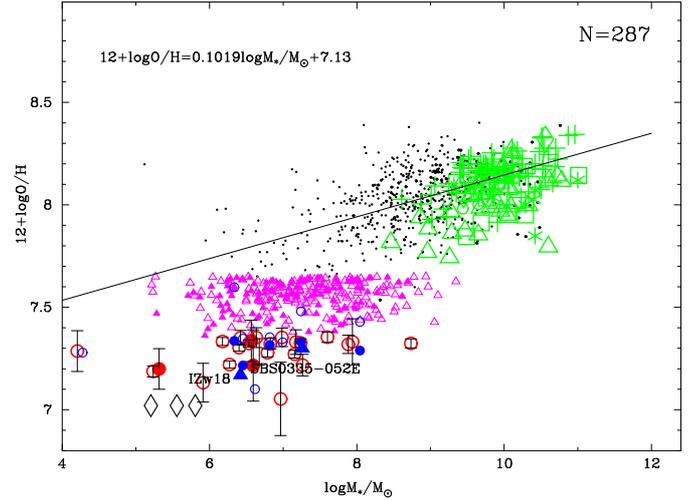}
\caption{Stellar mass - oxygen abundance relation for the same samples as 
in Fig. \ref{fig4} shown by the same symbols.
For comparison, the SDSS DR7 emission-line galaxy sample from \citet{I2014}, 
where the errors of the [O~{\sc iii}]$\lambda$4363 emission-line fluxes do not 
exceed 25\%, are shown by black dots with the linear likelihood regression 
for these data plotted by a solid black line.
The location of high-redshift star-forming galaxies is shown by various 
large green symbols (see text).
}
\label{fig5}
\end{figure}

\subsection{Luminosity-metallicity and mass-metallicity relations \label{S3s2}}

In this section, we discuss the distributions of the 
LMGs and XMDs in the 
 broad-band luminosity-metallicity, H$\beta$ luminosity-metallicity,
and mass-metallicity diagrams.
We compare our XMD candidates with similar galaxies analysed by
\citet{I06,I11,I12,I2014}, \citet{IT07}, and \citet{Guseva2015}. 

 The $L$ - $Z$ relation between optical 
luminosities expressed in extinction-corrected absolute SDSS $g$-band 
magnitudes $M_g$  (Table \ref{tab1})
and metallicities of galaxies
is shown in Fig. \ref{fig4}a for the LMG and XMD samples.
 The distances to the galaxies are derived from their radial velocities, 
adopting a Hubble constant $H_0$ = 67.3 km s$^{-1}$ Mpc$^{-1}$ \citep{Planck2014}.

   The position of all XMD candidates 
from this paper (large red filled and open circles) and from \citet{IT07}, 
\citet{I12}, and \citet{Guseva2015} (small blue circles)
are plotted.
 We also show the positions of the most metal-deficient BCDs --
I Zw 18, SBS 0335--052W, SBS 0335--052E (filled blue 
triangles) -- as well as Leo P 
and AGC 198691 (large open black square and diamonds in 
Fig. \ref{fig4}).
  For comparison, in Fig. \ref{fig4}a the 
SDSS galaxies by \citet{I06} and star-forming galaxies from \citet{IT07},
 as a representative sample, are plotted by black dots. 
 The standard $L$ - $Z$ relation for emission-line galaxies 
by \citet{I06} and \citet{IT07} is shown (solid line).
 We aim to examine how much the LMG and XMD samples
differ from the star-forming galaxies in standard samples.
  For the XMD sample we show error bars in oxygen abundances derived by the
 direct and semi-empirical methods.
  We note, that the errors in electron temperature, electron number density and
ionic and total abundances are usually larger for the direct method  because
the [O~{\sc iii}]$\lambda$4363 emission line is always weaker than 
[O~{\sc ii}]$\lambda$3727 and [O~{\sc iii}]$\lambda$5007 emission lines
used in so-called strong-line methods, particularly in the semi-empirical 
method. The direct method is considered to be the physically most 
reasonable method, while strong-line 
methods are subject to large systematic errors, despite the lower statistical 
errors.

  Most XMDs occupy the same region in the  $L - Z$ diagram as the well-known 
metal-deficient BCDs I Zw 18 and SBS 0335--052E. 
  All these galaxies are
systematically shifted to lower 
metallicities relative to the standard line for
every bin of the $M_g$ range from $-$12 to $-$18 mag.
  Considering a higher oxygen abundance bin 
7.35 $\leq$ 12 + log(O/H) $\leq$ 7.65 (magenta symbols)
we find a similar scatter of galaxies
in absolute magnitude. These galaxies fill the gap between the solid line 
in Fig. \ref{fig4}a and the location of XMDs with 12 + log(O/H) $\leq$ 7.35.
 The same conclusion was made by \citet{K04a} and \citet{G09,Guseva2015}.

  In Fig. \ref{fig4}b, the diagram of extinction- and aperture-corrected 
\citep[see ][]{I2014a} H$\beta$ luminosity 
$L$(H$\beta$)  metallicity is plotted for the same samples as in 
Fig. \ref{fig4}a. 
 The upper axis provides a scale for the star formation
rate (SFR), derived from the extinction- and aperture-corrected H$\alpha$ 
luminosity as defined by \citet{K98}.

  The XMD galaxies with oxygen abundances obtained by the direct method 
(three large red filled circles in Fig. \ref{fig4}) are shifted down to 
lower metallicities by about 0.4--0.6 dex at fixed $M_g$ and to brighter 
$M_g$ by more than 5--6 mag at fixed 12+logO/H relative to 
the standard $L-Z$ relation (black solid line).
 The galaxies with weaker or non-detected [O {\sc iii}]$\lambda$4363 emission 
line and hence with metallicities derived by the semi-empirical method 
are less shifted along the $M_g$ and $L$(H$\beta$) scales.

 As seen in Fig. \ref{fig4}b, most of our new XMD candidates
have H$\beta$ luminosities ranging from $\sim$10$^{38}$ to 
10$^{41}$ erg s$^{-1}$ and SFRs in the range 
$\sim$0.01 - 1.0 $M_\odot$ yr$^{-1}$, which are comparable
to the SFRs of I Zw 18 and the SBS 0335--052 system.
  Our XMDs with metallicities obtained using the semi-empirical 
method have a larger spread in $M_g$ and log $L$(H$\beta$) 
and about half of them have 
$L$(H$\beta$) in the range $\sim$10$^{35}$ to 10$^{38}$ erg s$^{-1}$
corresponding to lower (log SFR = -4-- -2 $M_\odot$ yr$^{-1}$) star formation 
activity in lower excitation H~{\sc ii} regions. They are located in the
region close to that of the low-mass galaxies in the leftmost end of the 
$L-Z$ and $L$(H$\beta$)-$Z$ relations (Fig. \ref{fig4}a and b, respectively) 
and which are populated by objects such as Leo P, AGC 198691 and J0143+1958 
\citep[][ respectively]{Skillman2013,Hirschauer2016,Guseva2015}.

  Summarising, intense starbursts with high SFRs, like the 
well-known BCDs I Zw 18 and SBS 0335--052E \citep{SS70,I90},
are present in 17 out of 23 XMD candidates. 
These XMD galaxies are relatively luminous and chemically unevolved galaxies 
with strong ongoing star formation. 
  Inclusion of these galaxies in
the $L-Z$ and $L$(H$\beta$)-$Z$ relations would result in shallower 
slopes \citep{G09}. 
 Furthermore, known XMDs with 12 + log(O/H) $<$ 7.35, 
together with the XMD candidates from this paper, do not show a 
dependence on metallicity.

  Stellar mass is one of the central parameters characterising galaxies, 
and the relation between this parameter and the galaxy 
metallicity is a fundamental relation between the two 
global galaxy properties. 
Mass is more fundamentally related to metallicity than luminosity.
 The luminosity of a galaxy, especially a dwarf galaxy, 
depends greatly on its present phase of star formation,
while stellar mass and metallicity are
both determined by the evolution of the galaxy.

To derive stellar masses we modelled spectral energy distributions (SEDs) 
in the entire spectral range of $\lambda$$\lambda$3600-10300$\AA$, 
subtracted nebular line and continuum emission which is significant 
in the spectra of star-forming galaxies.
  Neglecting the correction
for nebular continuum emission would result in a considerable overestimate of
the galaxy stellar mass.
 The SED of the nebular continuum is
taken from \citet{A84} and includes hydrogen and helium free-bound,
free-free, and two-photon emission.
  The shape of the spectrum also depends on reddening. 
The extinction coefficient  $C$(H$\beta$)
is obtained from the observed hydrogen Balmer decrement.
   To account for the contribution of the stellar emission, we have
adopted the grid of the Padua stellar evolution models by \citet{Gi00}.

   Each SED also depends on the adopted star formation history.
   The bursting nature of star formation and young burst ages have recently been
confirmed for a large sample of compact star-forming galaxies 
by \citet{Iz2016}.  
 The star formation in our galaxies occurs in short
strong bursts while the contribution of the continuous star formation is low.
So, we approximated the star formation history in each galaxy by a recent 
short burst with age $t$(young) $<$ 10 Myr, which accounts 
for the young stellar population, and a prior continuous star formation 
responsible for the older stars with age varying between 10 Myr and 15 Gyr.
The contribution of each stellar population to the SED is parameterised by 
the ratio of the masses of the old to young stellar populations, 
$b$ = $M$(old)/$M$(young), which we vary between 0.01 and 1000. 

  We carried out 10$^4$ Monte Carlo simulations to
reproduce the SED of each galaxy in our sample.
  The method is based on fitting the model SEDs to the observed ones and 
finding the best fit \citep[for details see e.g., ][]{G06,G07,I11,I2014a}. 

  In Fig. \ref{fig5} we show our 
LMG and XMD candidates on the extinction- and aperture-corrected
stellar mass-metallicity ($M_\star$-$Z$) diagram.
The same samples and symbols as in Fig. \ref{fig4} 
are used. The solid line is the $M_\star$-$Z$
relation obtained by \citet{I2014} for SDSS DR7 emission-line galaxies  
with robust metallicity determinations (black dots).

\setcounter{figure}{5}

\begin{figure*}
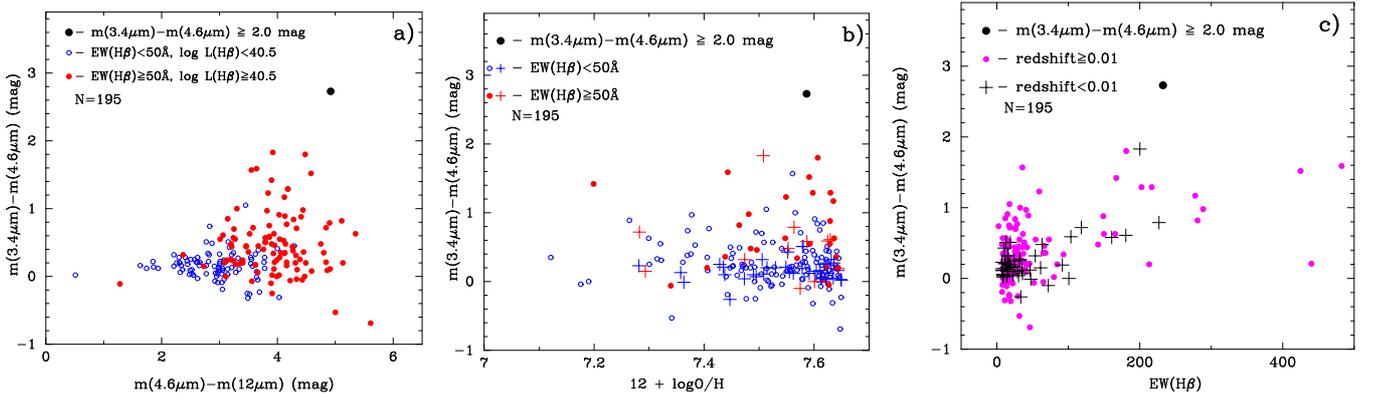

\hbox{
\hspace*{0.2cm}\psfig{figure=fig6a.ps,angle=-90,width=5.5cm,clip=}
\hspace*{0.2cm}\psfig{figure=fig6b.ps,angle=-90,width=5.6cm,clip=}
\hspace*{0.2cm}\psfig{figure=fig6c.ps,angle=-90,width=6.1cm,clip=}
}
\caption{ {\bf (a)} $m$(3.4$\mu$m) -- $m$(4.6$\mu$m) vs. 
$m$(4.6$\mu$m) -- $m$(12$\mu$m) colour-colour diagram
for a sample of emission-line galaxies detected in the three {\sl WISE} 
bands 3.4$\mu$m ($W1$), 4.6$\mu$m ($W2$), and 12$\mu$m ($W3$). 
Galaxies with equivalent widths of the H$\beta$ emission line 
EW(H$\beta$) $\geq$ 50 $\AA$\ and $L$(H$\beta$) $\geq$ 3 $\times$ 10$^{40}$ erg s$^{-1}$ are shown by filled red circles; galaxies with 
EW(H$\beta$) $<$ 50 $\AA$\ and 
$L$(H$\beta$) $<$ 3 $\times$ 10$^{40}$ erg s$^{-1}$
are depicted by open blue circles. 
  The galaxy with $m$(3.4$\mu$m) -- $m$(4.6$\mu$m) 
$=$ 2.73 mag (J1353+1649) discovered here, is shown by a large black filled 
circle in all three panels. 
{\bf (b)} $W$1 - $W$2 colour vs. oxygen abundance. 
The sample is split into two parts with high and low
EW(H$\beta$). Nearby galaxies with redshifts less than 0.01 are presented 
by crosses (red for the galaxies with high EW(H$\beta$)  
and blue for low equivalent widths of H$\beta$). The remaining galaxies are 
presented by the same symbols as in (a).
{\bf (c)} $W$1 - $W$2 vs. EW(H$\beta$) is shown for nearby galaxies (redshift 
$z$ $<$ 0.01, black crosses) 
and for more distant galaxies ($z$ $\geq$ 0.01, filled magenta circles).
}
\label{fig6}
\end{figure*}

   The 23 newly discovered XMD candidates are plotted by large red circles,
filled and open for direct and semi-empirical methods, respectively. 
 The XMDs are dwarf galaxies and their masses spread over three orders
of magnitude from 10$^5$ to 10$^8$$M_{\odot}$.
Most of them, except several low-mass galaxies with 
$M_\star$ $<$ 10$^6$ $M_\odot$, are located in the region of  
I Zw 18 and SBS 0335--052E.  
 For their range of stellar masses  
the oxygen abundances in these galaxies are $\sim$0.5 dex (or 3-4 times) 
lower than those predicted from  the $M_\star$-$Z$ relation for the 
SDSS DR7 emission-line galaxy sample.
   Thus, these galaxies are the least chemically evolved galaxies.
  Our low-metallicity sample (7.35 $\leq$ 12 + log O/H $\leq$ 7.65) 
denoted as magenta filled and open 
triangles in Fig. \ref{fig5} shows a similar spread in stellar masses 
spanning three orders of magnitude,
but are shifted to slightly larger masses 10$^6$ - 10$^9$$M_{\odot}$.
  Thus, some hint of the dependence of the stellar mass on metallicity 
is present if only XMD and LMG samples are considered.
However, this relation deviates from the standard 
$M_\star$-$Z$ relation produced for a
large non-metallicity-restricted sample of galaxies
\citep[see e.g., the sample with metallicities derived by the direct 
method by ][]{I2014a}.
For comparison, several samples of high-redshift star-forming galaxies
are shown in Fig. \ref{fig5} by various large green symbols.
  Specifically 6 galaxies are shown at $z$ $>$ 2 by \citet{Cullen2014}, 
96 galaxies at 2.0 $<$ $z$ $<$ 2.6 by \citet{Steidel2014}, 
6 galaxies at $z$ $\sim$ 2 by \citet{Maier2014},
7 galaxies at $z$ $\sim$ 3.5 by \citet{Maiolino2008},
30 galaxies at redshifts $z$ $\sim$ 3.4 by \citet{Toncosor2014}, 
and 4 galaxies at $z$ $\sim$ 2.3 by \citet{Sanders2015}.
Oxygen abundances 
for all high-$z$ galaxies are calculated with O3N2 or O3O2 calibrations 
from \citet{I2014a} (their equations 2 and 3).
A small shift to lower metallicities is present for these high-redshift 
galaxies compared to the standard relation for a large  
non-metallicity-restricted sample of local star-forming galaxies (black 
linear regression).

 With newly discovered XMDs we confirm our previous findings that the 
low-metallicity dwarf galaxies are characterised by very high specific star 
formation rates (sSFR) of $\sim$50 Gyr$^{-1}$, which are in the range 
of sSFRs for star-forming galaxies at high redshifts $\sim$2--4 
\citep{Cullen2014,Steidel2014,Maiolino2008,Maier2014,Toncosor2014}.

\subsection{{\sl WISE}  data for the sample of metal-poor candidates \label{S3s3}}

 In this section we study the properties of LMG and XMD candidates in
the mid-infrared range. To this end, the
SDSS spectroscopic data of our LMG sample were supplemented
by photometric data from the Wide-field Infrared Survey 
Explorer ({\sl WISE})  All-Sky Source Catalogue (ASSC) \citep{Wright2010}.
  Specifically, we use the mid-infrared photometry in $W$1, $W$2, 
and $W$3 bands ($m$(3.4$\mu$m),  $m$(4.6$\mu$m),  and $m$(12$\mu$m)), 
measured by the {\sl WISE} collaboration.
The typical errors in each band are less than $\pm$0.1 mag. 

One of our aims is to search for galaxies with warm and hot dust. 
  Galaxies with hot dust emission 
have been searched using the galaxies with
very red mid-infrared 
$m$(3.4$\mu$m) -- $m$(4.6$\mu$m) colours of more than 2 mag. 
  Very red $W$1 - $W$2 colour indicates
a sharp increase in intensity at 4.6$\mu$m as compared to 3.4$\mu$m owing to
the presence of hot dust in the galaxy. 
  This is opposite to the galaxies 
without hot dust emission which are characterised by bluer $W$1 - $W$2 
colours. More specifically \citet{I2014} showed that stellar and interstellar 
ionised gas emission are characterised by $W$1 - $W$2 colours of $\sim$0 and 
$\sim$0.7 mag, respectively.  
 This indicates that the colours $W$1 - $W$2 $\ga$ 2.0 mag can
be signatures of hot dust with temperatures reaching several hundred Kelvin.
 Such galaxies are fairly rare. Only 
7 galaxies were found in the {\sl WISE} 
Preliminary Release Source Catalogue (PRSC) 
by \citet{Griffith2011} and \citet{I11b} and 20 SDSS galaxies were found by
\citet{I2014} in the {\sl WISE} ASSC. 
 
  From the LMG sample 
we selected 195 galaxies, which were detected in the {\sl WISE} $W$1, $W$2 
and $W$3  bands.
These galaxies are shown in Fig. \ref{fig6}a on the ($W$1 - $W$2) vs. 
($W$2 - $W$3) colour-colour diagram. We split the sample
into two subsamples with equivalent widths of the H$\beta$ line 
EW(H$\beta$) $\geq$ 50 $\AA$\ and with the H$\beta$ luminosities 
$L$(H$\beta$) $\geq$ 3 $\times$ 10$^{40}$ erg s$^{-1}$ 
(red symbols) and EW(H$\beta$) $<$ 50 $\AA$\ and 
$L$(H$\beta$) $<$ 3 $\times$ 10$^{40}$ erg s$^{-1}$
(blue symbols). 
In the starburst scenario, which is 
relevant for our galaxies with strong ongoing star formation,
 EW(H$\beta$) is rapidly 
decreasing with age and therefore it is a good
signature of 
the star formation burst age. 

  Similarly to previous studies \citep{Griffith2011,I11b,I2014}, we 
did not find an appreciable number of new galaxies with very hot dust emission.
 The bulk of the galaxies ($\sim$85\%) have low mid-infrared colours
$W$1 - $W$2 $<$ 0.7 mag. There are 14 out 
of 195 galaxies with a considerable contribution of the hot dust emission and
$W$1 - $W$2 colours redder than 1--2 mag. 
  The very hot dust emission is likely
present only in one new galaxy J1353+1649 with oxygen abundance 
12 + logO/H $\sim$ 7.6 according to very red colour $W$1 - $W$2 of 2.73 mag 
(large black filled circle). 
 In general, the mid-infrared colour-colour diagrams of our metallicity-limited
sample (Fig. \ref{fig6}a) and of a large sample of $\sim$10 000 galaxies from 
the SDSS DR7 in \citet{I2014} (their Fig.7c) are very similar.
Most H$\beta$ luminous galaxies with high-excitation H {\sc ii} regions and 
smaller star formation ages have redder $W$1 - $W$2 and $W$2 - $W$3 colours 
(red points are clearly shifted to redder colours relative to the blue 
ones in Fig. \ref{fig6}a). 
  This implies that the main source of dust heating 
is radiation from young star-forming regions. 

 In Fig. \ref{fig6}b  our sample is split into two parts: 
EW(H$\beta$) $\geq$ 50 $\AA$\ (red filled circles) and 
EW(H$\beta$) $<$ 50 $\AA$\ (blue open circles). The nearby galaxies 
with redshifts less than 0.01 are presented in the figure
by crosses (red for the galaxies with high equivalent widths of H$\beta$ 
and blue for low EW(H$\beta$)).
   We note that the distant galaxies
with redshifts $z$ $>$ 0.01 in Fig. \ref{fig6}b have a much wider spread in 
colours than nearby galaxies. 
We have significantly added to the previously 
known galaxies in the range of 12 + logO/H from 7.2 to 7.65 with known
$W$1 - $W$2 colours as compared to the data by \citet{I11b,I2014} and 
\citet{Griffith2011} where the metallicity range is substantially wider. 
However, there is no dependence of mid-infrared colour on the metallicity 
for all these samples, including our.  Thus, metallicity is not 
the main factor affecting the heating of interstellar dust.

 It can be clearly seen in Fig. \ref{fig6}c that the $W$1 - $W$2 colour becomes
increasingly redder with increasing equivalent width of the H$\beta$ emission
line (larger EW(H$\beta$), corresponding to younger bursts of star formation) 
for both the nearby and more distant galaxies
shown by crosses and filled circles, respectively.
This suggests that the main contributor to the interstellar dust heating is the 
most recent burst of star formation.

\section{Conclusions \label{S4}}

We present a spectroscopic study of low-metallicity galaxy (LMG) 
emission-line candidates and extremely metal-deficient (XMD) 
emission-line candidates 
selected from the Data Release 12 (DR12) of the Sloan Digital Sky Survey (SDSS),
a part of the SDSS-III survey. 
 Our main results are as follows.
 
 We derived oxygen abundances of 12 + logO/H $\leq$ 7.65 in 287 
low-metallicity galaxies. Out of this number, the 
[O {\sc iii}]$\lambda$4363$\AA$\ 
emission line was detected at a level higher than 3$\sigma$ in 120 galaxies 
including three galaxies which were classified as XMD candidates.
This allowed us to derive their oxygen abundance using the direct method.
The semi-empirical method was used for the remaining 168 galaxies.

 Seventeen out of 23 newly discovered XMD candidates are 
characterised by strong SF activity and high SFRs, like two
well-known BCDs I Zw 18 and SBS 0335--052E \citep{SS70,I90}. 

 No emission-line galaxies with 12 + logO/H $\la$ 7.0 were found
in the entire SDSS DR12 survey and 
among the other galaxies collected from the literature 
\citep[this paper and ][]{I12,Guseva2015}.
This finding supports conclusions  
by \citet{T05}, \citet{I12}, and \citet{Guseva2015} that the matter 
from which XMDs were formed
was probably pre-enriched to this level prior to the galaxy formation. 

 The XMDs with strong SF activity are the most prominent outliers
in the luminosity-metallicity ($L-Z$), H$\beta$ luminosity-metallicity 
($L$(H$\beta$) - $Z$) and mass-metallicity relations, substantially
increasing the dispersion at the low-metallicity end of the relations.

 Extreme positions of the most massive and luminous XMDs in
the mass-metallicity diagrams imply that these galaxies with stellar masses  
$\sim$ 10$^7$ 
are not yet chemically enriched.

\begin{acknowledgements}

 Y.I.I. and N.G.G. thank the Max-Planck 
Institute for Radioastronomy, Bonn, Germany, for the hospitality.    
    Funding for the Sloan Digital Sky Survey (SDSS-III) has been 
provided by the Alfred P. Sloan Foundation, the Participating Institutions, 
the National Science Foundation, the U.S. Department of Energy, the National 
Aeronautics and Space Administration, the Japanese Monbukagakusho, and the 
Max Planck Society, the Higher Education Funding Council for England. 
 
\end{acknowledgements}



\setcounter{table}{1}

\begin{table*}
  \caption{Emission line intensities and equivalent widths\label{tab2}}
\begin{tabular}{lrrrcrrr} \hline \hline
 &\multicolumn{7}{c}{Galaxy} \\  \cline{2-8}       
 & \multicolumn{3}{c}{          J0122+0048}&&
 \multicolumn{3}{c}{          J0137+1810} \\ 
  \cline{2-4} \cline{6-8} 
  {Ion}
  &{$F$($\lambda$)/$F$(H$\beta$)}
  &{$I$($\lambda$)/$I$(H$\beta$)}
  &{EW$^{a}$}&
  &{$F$($\lambda$)/$F$(H$\beta$)}
  &{$I$($\lambda$)/$I$(H$\beta$)}
  &{EW$^{a}$} \\ \hline

3727 [O {\sc ii}]                 &  48.73 $\pm$   ~5.16 &  57.30 $\pm$   ~6.33 &  29.0 & &  27.51 $\pm$   ~7.74 &  34.01 $\pm$   ~9.90 &  35.7 \\
3869 [Ne {\sc iii}]               &  21.50 $\pm$   ~3.17 &  24.71 $\pm$   ~3.72 &  18.3 & &  28.27 $\pm$   ~7.24 &  33.89 $\pm$   ~8.95 &  31.0 \\
3889 He {\sc i} + H8              &  23.73 $\pm$   ~3.30 &  27.35 $\pm$   ~5.37 &  24.2 & &  17.78 $\pm$   ~5.53 &  25.02 $\pm$   ~8.89 &  23.7 \\
3968 [Ne {\sc iii}] + H7          &  25.41 $\pm$   ~3.53 &  28.89 $\pm$   ~5.05 &  31.9 & &  21.80 $\pm$   ~6.38 &  29.70 $\pm$   ~9.85 &  26.5 \\
4101 H$\delta$                    &  25.84 $\pm$   ~3.48 &  28.78 $\pm$   ~4.87 &  33.0 & &  27.25 $\pm$   ~7.59 &  35.26 $\pm$10.87 &  32.0 \\
4340 H$\gamma$                    &  45.72 $\pm$   ~5.02 &  49.05 $\pm$   ~5.82 &  78.5 & &  37.34 $\pm$   ~8.80 &  44.70 $\pm$11.48 &  43.1 \\
4363 [O {\sc iii}]                &   9.53 $\pm$   ~2.43 &  10.17 $\pm$   ~2.61 &  10.3 & &  17.07 $\pm$   ~5.83 &  18.52 $\pm$   ~6.43 &  34.7 \\
4861 H$\beta$                     & 100.00 $\pm$   ~8.74 & 100.00 $\pm$   ~9.06 & 147.5 & & 100.00 $\pm$17.13 & 100.00 $\pm$17.37 & 549.3 \\
4959 [O {\sc iii}]                &  86.73 $\pm$   ~7.86 &  85.60 $\pm$   ~7.80 & 179.6 & & 144.93 $\pm$23.02 & 141.49 $\pm$22.77 & 303.7 \\
5007 [O {\sc iii}]                & 245.77 $\pm$18.05 & 241.15 $\pm$17.81 & 468.7 & & 468.06 $\pm$62.76 & 453.37 $\pm$61.62 & 773.9 \\
5876 He {\sc i}                   &  11.18 $\pm$   ~2.70 &   9.97 $\pm$   ~2.42 &  22.7 & &  18.14 $\pm$   ~6.26 &  15.45 $\pm$   ~5.43 & 112.5 \\
6300 [O {\sc i}]                  &   1.39 $\pm$   ~1.31 &   1.20 $\pm$  ~1.13 &   3.2 & &...~~~~~~~~ &...~~~~~~~~ &   ... \\
6312 [S {\sc iii}]                &   0.63 $\pm$   ~1.21 &   0.54 $\pm$   ~1.04 &   1.5 & &...~~~~~~~~ &...~~~~~~~~ &   ... \\
6563 H$\alpha$                    & 324.53 $\pm$23.80 & 271.96 $\pm$21.77 &1015.0 & & 347.07 $\pm$49.60 & 272.49 $\pm$42.81 &1518.0 \\
6716 [S {\sc ii}]                 &   4.02 $\pm$   ~2.00 &   3.33 $\pm$   ~1.67 &   9.4 & &   8.46 $\pm$   ~4.95 &   6.51 $\pm$   ~3.88 &  29.9 \\
6731 [S {\sc ii}]                 &   3.76 $\pm$   ~1.87 &   3.11 $\pm$   ~1.55 &   9.0 & &   4.53 $\pm$   ~3.76 &   3.48 $\pm$   ~2.94 &  16.7 \\
9069 [S {\sc iii}]                &   1.75 $\pm$   ~1.32 &   1.26 $\pm$   ~0.99 &  17.9 & &...~~~~~~~~ &...~~~~~~~~ &   ... \\
9531 [S {\sc iii}]                &  20.74 $\pm$   ~3.76 &  14.52 $\pm$   ~2.80 &  56.6 & &...~~~~~~~~ &...~~~~~~~~ &   ... \\
 $C$(H$\beta$) & \multicolumn{3}{c}{ 0.225 }& & \multicolumn{3}{c}{ 0.320 } \\
 $F$(H$\beta$)$^{b}$ & \multicolumn{3}{c}{  8.10 }& & \multicolumn{3}{c}{  1.96 } \\
 EW(abs) $\AA$ & \multicolumn{3}{c}{ 0.41 }& & \multicolumn{3}{c}{ 2.41 } \\ 
\end{tabular}
\end{table*}

\setcounter{table}{1}

\begin{table*}
\begin{tabular}{lrrrcrrr} \hline
 &\multicolumn{7}{c}{Galaxy} \\  \cline{2-8} 
&\multicolumn{3}{c}{          J0141+2124}&&
 \multicolumn{3}{c}{          J0153+0104} \\
  \cline{2-4} \cline{6-8} 
  {Ion}
  &{$F$($\lambda$)/$F$(H$\beta$)}
  &{$I$($\lambda$)/$I$(H$\beta$)}
  &{EW$^{a}$}&
  &{$F$($\lambda$)/$F$(H$\beta$)}
  &{$I$($\lambda$)/$I$(H$\beta$)}
  &{EW$^{a}$} \\ \hline

3727 [O {\sc ii}]                 &  36.18 $\pm$11.64 &  42.71 $\pm$14.11 &  21.4 & &  22.04 $\pm$   ~2.92 &  24.21 $\pm$   ~3.29 &  34.1 \\
3869 [Ne {\sc iii}]               &  39.96 $\pm$11.43 &  46.09 $\pm$13.53 &  15.7 & &  37.97 $\pm$   ~3.96 &  41.14 $\pm$   ~4.42 &  47.0 \\
3889 He {\sc i} + H8              &  18.97 $\pm$   ~7.79 &  22.09 $\pm$12.48 &   7.6 & &  20.10 $\pm$   ~2.81 &  22.13 $\pm$   ~5.09 &  24.6 \\
3968 [Ne {\sc iii}] + H7          &  28.10 $\pm$   ~9.67 &  32.17 $\pm$13.91 &  11.5 & &  29.61 $\pm$   ~3.44 &  32.08 $\pm$   ~4.78 &  49.8 \\
4101 H$\delta$                    &  25.81 $\pm$   ~9.54 &  28.81 $\pm$11.23 &  26.7 & &  27.08 $\pm$   ~3.31 &  29.00 $\pm$   ~4.50 &  48.3 \\
4340 H$\gamma$                    &  49.87 $\pm$12.90 &  53.60 $\pm$14.31 &  54.6 & &  46.60 $\pm$   ~4.60 &  48.66 $\pm$   ~5.35 &  98.2 \\
4363 [O {\sc iii}]                &  16.95 $\pm$   ~7.88 &  18.12 $\pm$   ~8.48 &  14.9 & &  14.18 $\pm$   ~2.32 &  14.71 $\pm$   ~2.42 &  31.2 \\
4861 H$\beta$                     & 100.00 $\pm$20.15 & 100.00 $\pm$20.33 & 153.3 & & 100.00 $\pm$   ~8.06 & 100.00 $\pm$   ~8.36 & 215.5 \\
4959 [O {\sc iii}]                & 142.44 $\pm$26.65 & 140.60 $\pm$26.42 & 185.1 & & 132.51 $\pm$10.12 & 131.30 $\pm$10.07 & 180.7 \\
5007 [O {\sc iii}]                & 438.40 $\pm$69.22 & 430.15 $\pm$68.28 & 539.6 & & 401.86 $\pm$25.84 & 396.82 $\pm$25.64 & 582.4 \\
5876 He {\sc i}                   &  11.73 $\pm$   ~6.73 &  10.45 $\pm$   ~6.04 &  14.6 & &   9.68 $\pm$   ~2.24 &   9.03 $\pm$   ~2.11 &  41.6 \\
6563 H$\alpha$                    & 325.49 $\pm$54.69 & 271.82 $\pm$49.79 & 745.7 & & 303.05 $\pm$20.78 & 272.84 $\pm$20.41 &1432.0 \\
6583 [N {\sc ii}]                 &  17.14 $\pm$   ~7.95 &  14.29 $\pm$   ~6.73 &  46.7 & &...~~~~~~~~ &...~~~~~~~~ &   ... \\
6716 [S {\sc ii}]                 &...~~~~~~~~ &...~~~~~~~~ &   ... & &   1.82 $\pm$   ~1.20 &   1.62 $\pm$   ~1.08 &  10.0 \\
6731 [S {\sc ii}]                 &...~~~~~~~~ &...~~~~~~~~ &   ... & &   1.84 $\pm$   ~1.21 &   1.64 $\pm$   ~1.09 &   9.2 \\
9069 [S {\sc iii}]                &...~~~~~~~~ &...~~~~~~~~ &   ... & &   4.13 $\pm$   ~1.84 &   3.42 $\pm$   ~1.60 &  38.2 \\
 $C$(H$\beta$) & \multicolumn{3}{c}{ 0.240 }& & \multicolumn{3}{c}{ 0.135 } \\
 $F$(H$\beta$)$^{b}$ & \multicolumn{3}{c}{  1.57 }& & \multicolumn{3}{c}{  9.35 } \\
 EW(abs) $\AA$ & \multicolumn{3}{c}{ 0.00 }& & \multicolumn{3}{c}{ 0.75 } \\ \hline
\end{tabular}

\end{table*}

\setcounter{table}{1}

\begin{table*}
\begin{tabular}{lrrrcrrr} \hline
 &\multicolumn{7}{c}{Galaxy} \\  \cline{2-8} 
&\multicolumn{3}{c}{          J0739+4434}&&
 \multicolumn{3}{c}{          J0945+3835} \\ 
  \cline{2-4} \cline{6-8} 
  {Ion}
  &{$F$($\lambda$)/$F$(H$\beta$)}
  &{$I$($\lambda$)/$I$(H$\beta$)}
  &{EW$^{a}$}&
  &{$F$($\lambda$)/$F$(H$\beta$)}
  &{$I$($\lambda$)/$I$(H$\beta$)}
  &{EW$^{a}$} \\ \hline

3727 [O {\sc ii}]                 &  70.88 $\pm$13.74 &  69.43 $\pm$14.49 &  62.6 & & 103.09 $\pm$12.51 & 129.57 $\pm$16.80 &  41.2 \\
3835 H9                           &  26.24 $\pm$  ~6.77 &  25.98 $\pm$  ~8.67 & 300.0 & &...~~~~~~~~ &...~~~~~~~~ &   ... \\
3869 [Ne {\sc iii}]               &  51.05 $\pm$11.13 &  50.00 $\pm$11.49 &  44.1 & &  34.98 $\pm$  ~6.04 &  42.57 $\pm$  ~7.58 &  19.7 \\
3889 He {\sc i} + H8              &...~~~~~~~~ &...~~~~~~~~ &   ... & &  32.60 $\pm$  ~5.48 &  39.60 $\pm$  ~7.22 &  57.2 \\
3968 [Ne {\sc iii}] + H7          &...~~~~~~~~ &...~~~~~~~~ &   ... & &  31.17 $\pm$  ~5.85 &  37.35 $\pm$  ~8.92 &  22.9 \\
4101 H$\delta$                    &  23.09 $\pm$  ~7.27 &  26.38 $\pm$70.49 &  19.2 & &  23.91 $\pm$  ~5.08 &  27.82 $\pm$  ~7.01 &  25.0 \\
4340 H$\gamma$                    &  45.58 $\pm$11.14 &  47.88 $\pm$61.27 &  44.2 & &  45.88 $\pm$  ~7.40 &  50.71 $\pm$  ~9.15 &  42.8 \\
4363 [O {\sc iii}]                &  14.36 $\pm$  ~6.14 &  14.07 $\pm$  ~6.16 &  12.2 & &   8.94 $\pm$  ~3.66 &   9.80 $\pm$  ~4.04 &   9.4 \\
4686 He {\sc ii}                  &...~~~~~~~~ &...~~~~~~~~ &   ... & &   6.30 $\pm$  ~3.04 &   6.50 $\pm$  ~3.16 &   4.6 \\
4861 H$\beta$                     & 100.00 $\pm$18.26 & 100.00 $\pm$42.44 & 152.8 & & 100.00 $\pm$12.72 & 100.00 $\pm$13.34 &  87.6 \\
4959 [O {\sc iii}]                & 111.88 $\pm$19.87 & 109.58 $\pm$19.87 & 192.2 & &  74.28 $\pm$10.19 &  72.90 $\pm$10.07 &  79.4 \\
5007 [O {\sc iii}]                & 367.07 $\pm$53.42 & 359.54 $\pm$53.46 &1113.0 & & 248.84 $\pm$26.40 & 242.20 $\pm$25.89 & 246.3 \\
5876 He {\sc i}                   &...~~~~~~~~ &...~~~~~~~~ &   ... & &  14.47 $\pm$  ~4.11 &  12.31 $\pm$  ~3.54 &  27.2 \\
6563 H$\alpha$                    & 264.03 $\pm$41.11 & 260.11 $\pm$52.70 & 556.1 & & 350.10 $\pm$36.71 & 272.69 $\pm$31.31 & 440.3 \\
 $C$(H$\beta$) & \multicolumn{3}{c}{ 0.000 }& & \multicolumn{3}{c}{ 0.330 } \\
 $F$(H$\beta$)$^{b}$ & \multicolumn{3}{c}{  1.81 }& & \multicolumn{3}{c}{  3.98 } \\
 EW(abs) $\AA$ & \multicolumn{3}{c}{ 2.00 }& & \multicolumn{3}{c}{ 0.22 } \\ 
\end{tabular}

\end{table*}

\setcounter{table}{1}

\begin{table*}
\begin{tabular}{lrrrcrrr} \hline
 &\multicolumn{7}{c}{Galaxy} \\  \cline{2-8}
& \multicolumn{3}{c}{          J1220+4915}&&
 \multicolumn{3}{c}{          J1444+4237} \\ 
  \cline{2-4} \cline{6-8} 
{Ion}
  &{$F$($\lambda$)/$F$(H$\beta$)}
  &{$I$($\lambda$)/$I$(H$\beta$)}
  &{EW$^{a}$}&
  &{$F$($\lambda$)/$F$(H$\beta$)}
  &{$I$($\lambda$)/$I$(H$\beta$)}
  &{EW$^{a}$} \\ \hline

3727 [O {\sc ii}]                 &  21.72 $\pm$  ~2.50 &  22.31 $\pm$  ~2.62 &  15.0 & & ...~~~~~~~~ &...~~~~~~~~ &   ...                  \\
3835 H9                           &   8.45 $\pm$  ~1.45 &   8.85 $\pm$  ~2.04 &  11.2 & &  8.77 $\pm$  ~1.95 &  10.41 $\pm$  ~3.07 &   2.5 \\
3869 [Ne {\sc iii}]               &  29.24 $\pm$  ~2.79 &  29.92 $\pm$  ~2.93 &  33.2 & & 17.65 $\pm$  ~2.59 &  19.00 $\pm$  ~2.85 &   2.5\\
3889 He {\sc i} + H8              &  18.77 $\pm$  ~2.20 &  19.44 $\pm$  ~2.88 &  19.4 & & 22.69 $\pm$  ~2.62 &  25.37 $\pm$  ~3.65 &   6.2 \\
3968 [Ne {\sc iii}] + H7          &  26.50 $\pm$  ~2.71 &  27.25 $\pm$  ~3.21 &  31.4 & & 15.36 $\pm$  ~2.84 &  17.91 $\pm$  ~4.66 &   2.7\\
4101 H$\delta$                    &  27.42 $\pm$  ~2.80 &  28.10 $\pm$  ~3.29 &  31.7 & & 24.39 $\pm$  ~2.97 &  26.93 $\pm$  ~4.22 &   5.3\\
4340 H$\gamma$                    &  49.58 $\pm$  ~3.97 &  50.30 $\pm$  ~4.28 &  64.4 & & 45.08 $\pm$  ~4.06 &  47.64 $\pm$  ~4.91 &  11.0\\
4363 [O {\sc iii}]                &  11.65 $\pm$  ~1.87 &  11.77 $\pm$  ~1.90 &  14.0 & & 8.17 $\pm$  ~2.61 &   8.43 $\pm$  ~2.73 &   1.5 \\
4471 He {\sc i}                   &   4.75 $\pm$  ~1.40 &   4.79 $\pm$  ~1.41 &   5.6 & &  3.43 $\pm$  ~2.26 &   3.50 $\pm$  ~2.33 &   0.7 \\
4713 [Ar {\sc iv}] + He {\sc i}   &   2.83 $\pm$  ~1.30 &   2.84 $\pm$  ~1.31 &   3.7 & &...~~~~~~~~ &...~~~~~~~~ &   ...                   \\
4861 H$\beta$                     & 100.00 $\pm$  ~6.63 & 100.00 $\pm$  ~6.73 & 167.1 & & 100.00 $\pm$  ~6.52 & 100.00 $\pm$  ~6.82 &  30.1\\
4959 [O {\sc iii}]                & 100.06 $\pm$  ~6.68 &  99.70 $\pm$  ~6.67 & 172.4 & & 83.26 $\pm$  ~5.88 &  81.99 $\pm$  ~5.85 &  22.5 \\
5007 [O {\sc iii}]                & 300.88 $\pm$16.45 & 299.50 $\pm$16.42 & 475.5 & & 243.96 $\pm$13.42 & 239.42 $\pm$13.31 &  66.3\\
5015 He {\sc i}                   &   2.75 $\pm$  ~1.18 &   2.73 $\pm$  ~1.18 &   4.5 & &  2.27 $\pm$  ~1.91 &   2.23 $\pm$  ~1.89 &   0.6 \\
5518 [Cl {\sc iii}]               &...~~~~~~~~ &...~~~~~~~~ &   ... & &   1.97 $\pm$  ~1.62 &   1.87 $\pm$  ~1.56 &   0.6 \\
5538 [Cl {\sc iii}]               &...~~~~~~~~ &...~~~~~~~~ &   ... & &   1.00 $\pm$  ~1.50 &   0.95 $\pm$  ~1.44 &   0.3 \\
5876 He {\sc i}                   &  13.02 $\pm$  ~2.04 &  12.74 $\pm$  ~2.01 &  49.2 & &   6.37 $\pm$  ~2.28 &   5.91 $\pm$  ~2.14 &   2.3\\
6300 [O {\sc i}]                  &...~~~~~~~~ &...~~~~~~~~ &   ... & &  3.61 $\pm$  ~2.34 &   3.27 $\pm$  ~2.14 &   1.4\\
6312 [S {\sc iii}]                &...~~~~~~~~ &...~~~~~~~~ &   ... & &  0.79 $\pm$  ~1.58 &   0.71 $\pm$  ~1.44 &   0.3\\
6563 H$\alpha$                    & 280.20 $\pm$16.02 & 271.44 $\pm$16.89 & 793.9 & & 305.55 $\pm$16.69 & 273.78 $\pm$16.43 & 146.5\\
6583 [N {\sc ii}]                 &...~~~~~~~~ &...~~~~~~~~ &   ... & &  6.12 $\pm$  ~2.42 &   5.47 $\pm$  ~2.19 &   2.9\\
6678 He {\sc i}                   &   2.49 $\pm$  ~1.35 &   2.40 $\pm$  ~1.31 &   8.0 & &  2.43 $\pm$  ~1.76 &   2.17 $\pm$  ~1.58 &   1.2 \\
6716 [S {\sc ii}]                 &...~~~~~~~~ &...~~~~~~~~ &   ...  & & 15.53 $\pm$  ~3.09 &  13.79 $\pm$  ~2.80 &   7.5 \\
6731 [S {\sc ii}]                 &...~~~~~~~~ &...~~~~~~~~ &   ... & & 11.85 $\pm$  ~3.20 &  10.51 $\pm$  ~2.88 &   5.8 \\
7065 He {\sc i}                   &   6.96 $\pm$  ~1.78 &   6.69 $\pm$  ~1.72 &  24.6 & &...~~~~~~~~ &...~~~~~~~~ &   ...\\
7136 [Ar {\sc iii}]               &...~~~~~~~~ &...~~~~~~~~ &   ... & &  2.89 $\pm$  ~2.00 &   2.52 $\pm$  ~1.76 &   1.5\\
7320 [O {\sc ii}]                 &...~~~~~~~~ &...~~~~~~~~ &   ... & &  2.69 $\pm$  ~1.73 &   2.32 $\pm$  ~1.51 &   1.7\\
7330 [O {\sc ii}]                 &...~~~~~~~~ &...~~~~~~~~ &   ... & &  1.33 $\pm$  ~1.58 &   1.15 $\pm$  ~1.38 &   0.8\\
9069 [S {\sc iii}]                &   2.79 $\pm$  ~1.34 &   2.67 $\pm$  ~1.33 &  15.3 & & 13.33 $\pm$  ~2.61 &  11.03 $\pm$  ~2.27 &  16.6\\
9531 [S {\sc iii}]                &   4.88 $\pm$  ~1.58 &   4.58 $\pm$  ~1.50 &  77.7 & &...~~~~~~~~ &...~~~~~~~~ &   ...\\
 $C$(H$\beta$) & \multicolumn{3}{c}{ 0.040 }& & \multicolumn{3}{c}{ 0.140 } \\
 $F$(H$\beta$)$^{b}$ & \multicolumn{3}{c}{ 14.27 }& & \multicolumn{3}{c}{ 17.83 } \\
 EW(abs) $\AA$ & \multicolumn{3}{c}{ 0.25 }& & \multicolumn{3}{c}{ 0.10 } \\ \hline
\end{tabular}

$^{\rm a}$In angstroms. \\
$^{\rm b}$In units of 10$^{-16}$erg s$^{-1}$cm$^{-2}$. 

\end{table*}

\setcounter{table}{2}

\begin{table*}
  \caption{Ionic and total heavy element abundances\label{tab3}}
  \begin{tabular}{lcccc} \hline \hline
&\multicolumn{4}{c}{Galaxy} \\  \cline{1-5} 
{Property}&
\multicolumn{1}{c}{        J0122+0048} & \multicolumn{1}{c}{        J0137+1810} &
\multicolumn{1}{c}{        J0141+2124}  &
\multicolumn{1}{c}{        J0153+0104}   \\
  \cline{1-5}
  $T_{\rm e}$(O {\sc iii}) (K) &
$22858.\pm 4181.$ & $22885.\pm 5791.$ &
$23202.\pm 7929.$  &
$21236.\pm 2367.$  \\

  $T_{\rm e}$(O {\sc ii}) (K) &
$15412.\pm 5568.$ & $15043.\pm 7715.$ &
$14973.\pm10622.$  &
$15384.\pm 3064.$  \\

  $N_{\rm e}$(S {\sc ii}) (cm$^{-3}$) &
494. & 100. &
100. &
694. \\

  \\
  \\
  O$^+$/H$^+$ ($\times$10$^4$) &
$ 0.050\pm 0.046$ & $ 0.031\pm 0.042$ &
$ 0.039\pm 0.073$  &
$ 0.022\pm 0.011$  \\

  O$^{++}$/H$^+$ ($\times$10$^4$) &
$ 0.108\pm 0.043$ & $ 0.196\pm 0.108$ &
$ 0.183\pm 0.136$  &
$ 0.200\pm 0.050$   \\

  O/H ($\times$10$^4$) &
$ 0.158\pm 0.063$ & $ 0.226\pm 0.116$ &
$ 0.222\pm 0.154$  &
$ 0.222\pm 0.052$   \\

  12 + log(O/H)  &
$ 7.199\pm 0.174$ & $ 7.354\pm 0.223$ &
$ 7.346\pm 0.302$  &
$ 7.347\pm 0.101$   \\

  \\
  Ne$^{++}$/H$^+$ ($\times$10$^5$) &
$ 0.233\pm 0.088$ & $ 0.319\pm 0.175$ &
$ 0.422\pm 0.297$  &
$ 0.455\pm 0.113$   \\

  ICF &
  1.135 &  1.057 &
  1.074  &
  1.042   \\

  log(Ne/O) &
$-0.776\pm 0.246$ & $-0.826\pm 0.342$ &
$-0.690\pm 0.450$  &
$-0.671\pm 0.156$   \\

\end{tabular}
\end{table*}

\setcounter{table}{2}

\begin{table*}
  \begin{tabular}{lcccc} \hline
&\multicolumn{4}{c}{Galaxy} \\  \cline{1-5}
{Property}&
\multicolumn{1}{c}{        J0739+4434} & \multicolumn{1}{c}{        J0945+3835} &
\multicolumn{1}{c}{        J1220+4915}  &
\multicolumn{1}{c}{        J1444+4237} \\ \hline

  $T_{\rm e}$(O {\sc iii}) (K) &
$22315.\pm 6867.$ &
$22873.\pm 6698.$ & 
$22061.\pm 2472.$  &
$20466.\pm 4152.$  \\

  $T_{\rm e}$(O {\sc ii}) (K) &
$15206.\pm 9058.$ &
$15084.\pm 8921.$ & 
$15547.\pm 3246.$  &
$15469.\pm 5305.$  \\

  $N_{\rm e}$(S {\sc ii}) (cm$^{-3}$) &
100. &
100. & 
100. &
111. \\
  \\
  \\
  O$^+$/H$^+$ ($\times$10$^4$) &
$ 0.060\pm 0.093$ &
$ 0.115\pm 0.178$ & 
$ 0.018\pm 0.010$  &
$ 0.079\pm 0.068$ \\
  O$^{++}$/H$^+$ ($\times$10$^4$) &
$ 0.162\pm 0.110$ &
$ 0.104\pm 0.066$ & 
$ 0.141\pm 0.035$  &
$ 0.131\pm 0.060$ \\
  O/H ($\times$10$^4$) &
$ 0.222\pm 0.144$ &
$ 0.219\pm 0.189$ & 
$ 0.159\pm 0.036$  &
$ 0.210\pm 0.091$ \\
  12 + log(O/H)  &
$ 7.347\pm 0.281$ &
$ 7.341\pm 0.375$ & 
$ 7.201\pm 0.099$  &
$ 7.323\pm 0.188$ \\
  \\
  Ne$^{++}$/H$^+$ ($\times$10$^5$) &
$ 0.497\pm 0.316$ &
$ 0.402\pm 0.234$ & 
$ 0.305\pm 0.073$  &
$ 0.228\pm 0.101$ \\
  ICF &
  1.115 &
  1.231 &  
  1.049  &
  1.160 \\
  log(Ne/O) &
$-0.603\pm 0.408$ &
$-0.647\pm 0.458$ & 
$-0.697\pm 0.151$  &
$-0.899\pm 0.275$ \\
   \hline
\end{tabular}
\end{table*}


\begin{thebibliography}{}
\bibitem[Alam et al.(2015)]{A15} Alam, S., et al., 
2015, \apjs, 219, 12 
\bibitem[Aller (1984)]{A84} Aller, L. H. 1984, Physics of Thermal Gaseous 
Nebulae (Dordrecht: Reidel)
\bibitem[Baldwin et al.(1981)]{BPT81} Baldwin, J. A., Phillips, M. M., \&
Terlevich, R. 1981, \pasp, 93, 5
\bibitem[Berg et al.(2012)]{B12} Berg, D. A., Skillman, E. D., Marble, A. R.,
et al. 2012, \apj, 754, 98
\bibitem[Cardelli et al. (1989)]{Cardelli1989} Cardelli J. A., Clayton G. C., 
Mathis J. S., 1989, \apj, 345, 245
\bibitem[Cullen et al. (2014)]{Cullen2014} Cullen, F., Cirasuolo, M., 
McLure, R. J., Dunlop, J.S., Bowler, R. A. A.,  2014, \mnras, 440, 2300
\bibitem[Fioc \& Rocca-Volmerange (1997)]{FR97} Fioc, M., \& 
Rocca-Volmerange, B. 1997, \aap, 326, 950
\bibitem[Girardi et al. (2000)]{Gi00} Girardi, L., Bressan, A., Bertelli, G., 
\& Chiosi, C. 2000, \aaps, 141, 371
\bibitem[Griffith et al. (2011)]{Griffith2011} Griffith, R.L., Tsai, C.-W., 
Stern, D., et al. 2011, \apj, 736, L22
\bibitem[Guseva et al.(2006)]{G06} Guseva, N. G., Izotov, Y. I., \&
Thuan, T. X. 2006, \apj, 644, 890
\bibitem[Guseva et al.(2007)]{G07} Guseva, N. G., Izotov, Y. I., 
Papaderos, P.  \& Fricke, K. J. 2007, \aap, 464, 885
\bibitem[Guseva et al.(2009)]{G09} Guseva, N. G., Papaderos, P., Meyer, H.,
Izotov, Y. I., \& Fricke, K. J. 2009, \aap, 505, 63
\bibitem[Guseva et al.(2011)]{Guseva2011}  Guseva, N. G., Izotov, Y. I.,
Stasi\'nska, G., Fricke, K. J., Henkel, C. \& Papaderos, P. 2011, \aap, 529, 149
\bibitem[Guseva et al.(2013)]{Guseva2013}  Guseva, N. G., Izotov, Y. I., 
Fricke, K. J. \& Henkel, C.  2013, \aap, 555, 90
\bibitem[Guseva et al.(2015)]{Guseva2015}  Guseva, N. G., Izotov, Y. I., 
Fricke, K. J. \& Henkel, C.  2015, \aap, 579, 11
\bibitem[Hirschauer et al.(2016)]{Hirschauer2016} Hirschauer, A. S., 
Salzer, J. J., Skillman, E. D., et al., 2016, \apj, 822, 108
\bibitem[Izotov \& Thuan(2007)]{IT07} Izotov, Y. I., \& Thuan, T. X. 2007, 
\apj, 665, 1115
\bibitem[Izotov \& Thuan(2009)]{IT09} Izotov, Y. I., \& Thuan, T. X. 2009, 
\apj, 690, 1797
\bibitem[Izotov et al.(1990)]{I90} Izotov, Y. I., Guseva, N. G., 
Lipovetsky, V. A., Kniazev, A. Y., \& Stepanian, J. A. 1990, \nat, 343, 238
\bibitem[Izotov et al.(1994)]{ITL94} Izotov, Y. I., Thuan, T. X., \& 
Lipovetsky, V. A. 1994, \apj, 435, 647 
\bibitem[Izotov et al.(1997)]{ITL97} Izotov, Y. I., Thuan, T. X., \& 
Lipovetsky, V. A. 1997, \apjs, 108, 1
\bibitem[Izotov et al.(2006a)]{I06} Izotov, Y. I., Stasi\'nska, G., 
Meynet, G., Guseva, N. G., \& Thuan, T. X. 2006a, \aap, 448, 955
\bibitem[Izotov et al.(2006b)]{I06b} Izotov, Y. I., Papaderos, P., 
Guseva, N. G., Fricke, K. J., \& Thuan, T. X. 2006b, \aap, 454, 137
\bibitem[Izotov et al.(2009)]{I09} Izotov, Y. I., Guseva, N. G.,
Fricke, K. J., \& Papaderos, P. 2009, \aap, 503, 61
\bibitem[Izotov et al.(2011)]{I11} Izotov, Y. I., Guseva, N. G., \&
Thuan, T. X. 2011, \apj, 728, 161
\bibitem[Izotov et al.(2011b)]{I11b} Izotov, Y. I., Guseva, N. G., Fricke, K. J.
\& Henkel, C. 2011b, \aap, 536, L7
\bibitem[Izotov et al.(2012)]{I12} Izotov, Y. I., Thuan, T. X. \& 
Guseva, N. G. 2012, \aap, 546, 122
\bibitem[Izotov et al.(2014a)]{I2014} Izotov, Y. I., Guseva, N. G., 
Fricke, K. J., \& Henkel, C. 2014a, \aap, 561, 33
\bibitem[Izotov et al.(2015)]{I2014a} Izotov, Y. I., Guseva, N. G., 
Fricke, K. J., \& Henkel, C. 2015, \mnras, 451, 2251
\bibitem[Izotov et al.(2016)]{Iz2016} Izotov, Y. I., Guseva, N. G., 
Fricke, K. J., \& Henkel, C. 2016, \mnras, 462, 4427
\bibitem[James et al.(2014)]{James2014} James, B. L., Koposov, S., Stark, D. P.
et al. 2014, MNRAS, arXiv: astro-ph 1411.7371.v1 
\bibitem[Kauffmann et al.(2003)]{K03} Kauffmann, G., Heckman, T. M., 
Tremonti, C., et al. 2003, \mnras, 346, 1055
\bibitem[Kennicutt(1998)]{K98} Kennicutt, R. C., Jr. 1998, \araa, 36, 189
\bibitem[Kniazev et al.(2003)]{K04a} Kniazev, A. Y., Grebel, E. K.,
Hao, L., Strauss, M. A., Brinkmann, J., \& Fukugita, M. 2003, \apj, 593, 73
\bibitem[Kunth \& \"Ostlin(2000)]{KO2000} Kunth, D., \& \"Ostlin, G. 2000,
\aap Rev., 10, 1
\bibitem[Maier et al.(2014)]{Maier2014} Maier, C., Lilly, S. J., 
Ziegler, B. L., P\'erez Montero, E., Peng, Y., \& Balestra, I. 2014, 
\apj, 792, 3
\bibitem[Maiolino et al.(2008)]{Maiolino2008} Maiolino, R, et al. 2008,
\aap, 488, 463
\bibitem[Perepelitsyna et al.(2014)]{Perepel2014} Perepelitsyna, Y. A., 
Pustilnik, S. A. Kniazev, A. Y. 2014, Astrophys. Bull., 69, 247
\bibitem[Planck collaboration XVI, (2014)]{Planck2014} Planck collaboration XVI,
2014, \aap, 571, A16
\bibitem[Prochaska et al.(2003)]{P03} Prochaska, J. X., Gawiser, E., Wolfe, A. M., Castro, S., \& Djorgovski, S. G. 2003, \apj, 595, L9
\bibitem[Pustilnik et al.(2005)]{P05} Pustilnik, S. A., Kniazev, A. Y., \&
Pramskij, A. G. 2005, \aap, 443, 91
\bibitem[Pustilnik et. al.(2010)]{Pustilnik2010} Pustilnik, S. A., 
Teplyakova, A. L., Kniazev, A. Y., Martin, J.-M., \& Burenkov, A. N. 2010,
\mnras, 401, 333
\bibitem[Pustilnik et. al.(2011)]{Pustilnik2011} Pustilnik, S. A., 
Teplyakova, A. L., Kniazev, A. Y. 2011,  Astrophys. Bull., 66, 255
\bibitem[Richer \& McCall(1995)]{RM95} Richer, M. G., 
\& McCall, M. L. 1995, \apj, 445, 642
\bibitem[S\'anchez Almeida et al.(2016)]{Almeida2016} S\'anchez Almeida, J.,
P\'erez-Montero, E., Morales-Luis, A. B., et al. 2016, \apj, 819, 110
\bibitem[Sanders et al.(2015)]{Sanders2015} Sanders, R. L., et al. 2015, 
\apj, 799, 138
\bibitem[Sargent \& Searle(1970)]{SS70} Sargent, W. L. W., \& Searle, L.
1970, \apj, 162, L155
\bibitem[Skillman et al.(1989)]{S89} Skillman, E. D., Kennicutt, R. C., Jr.
\& Hodge, P. W. 1989, \apj, 347, 875
\bibitem[Skillman et al.(2013)]{Skillman2013} Skillman, E. D., Salzer, J. J.,
Berg, D. A. et al, 2013, \aj, 146, 3
\bibitem[Stasi\'nska \& Izotov(2003)]{SI03} Stasi\'nska, G., \&
Izotov, Y. I. 2003, \aap, 397, 71
\bibitem[Stasi\'nska et al.(2006)]{S06} Stasi\'nska, G., Cid Fernandes, R., 
Mateus, A., Sodr\'e, L., \& Asari, N. V. 2006, \mnras, 371, 972
\bibitem[Steidel et al.(2014)]{Steidel2014} Steidel, , C. S., et al. 2014,
\apj, 795, 165 
\bibitem[Thuan \& Izotov(2005)]{TI2005} Thuan, T. X., \& Izotov, Y. I.
2005, \apjs, 161, 240
\bibitem[Thuan et al.(1995)]{TIL95} Thuan, T. X., 
Izotov, Y. I., \& Lipovetsky, V. A. 1995, \apj, 445, 108
\bibitem[Thuan et al.(2005)]{T05} Thuan, T. X., Lecavelier des Etangs, A.,
\& Izotov, Y. I. 2005, \apj, 621, 269
\bibitem[Tremonti et al.(2004)]{Tremonti2004} Tremonti, C., et al. 
2004, \apj, 613, 898
\bibitem[Troncoso et al.(2014)]{Toncosor2014} Troncoso, P., et al. 2014,
\aap, 563,A58
\bibitem[van Zee(2000)]{vZee2000} van Zee, L. 2000,\apj, 543, L31
\bibitem[Whitford(1958)]{W58} Whitford, A. E. 1958, \aj, 63, 201
\bibitem[Wright (2006)]{Wright2006} Wright, E. L. 2006, \pasp, 118, 1711
\bibitem[Wright et al.(2010)]{Wright2010} Wright, E. L., Eisenhardt, P. R. M., 
Mainzer, A. K., et al. 2010, \aj, 140, 1868


\end{thebibliography}
\end{document}